\documentclass[namedreferences]{solarphysics}

\usepackage[hyperref,optionalrh,showbiblabels]{spr-sola-addons} 
\usepackage{graphicx}        
\usepackage{xcolor}           
\hypersetup{
		 colorlinks=true,
		 linkcolor=blue,
		 filecolor=magenta,
		 urlcolor=blue,breaklinks
		 }
\usepackage{hyperref}
\usepackage{epstopdf}
 \usepackage{auto-pst-pdf}
\usepackage{url}

\newcommand{\etal}{{\it et al.}}
\newcommand{\ie}{{\it i.e. }}
\newcommand{\eg}{{\it e.g. }}
\newcommand{\ecc}{{\it etc. }}

\newcommand{\soho}{SOHO}

\newcommand{\adv}{    {\it Adv. Space Res.}} 
 
\newcommand{\aap}{    {\it Astron. Astrophys.}}

\newcommand{\apj}{    {\it Astrophys. J.}}
\newcommand{\apjl}{   {\it Astrophys. J. Lett.}}

\newcommand{\jgr}{    {\it J. Geophys. Res.}}

\newcommand{\solphys}{{\it Solar Phys.}}
 
\newcommand{\ssr}{    {\it Space Sci. Rev.}} 
\chardef\us=`\_
\ifx \doiurl    \undefined \def \doiurl#1{\href{http://dx.doi.org/#1}{\textsf{DOI}}}\fi
\ifx \adsurl    \undefined \def \adsurl#1{\href{http://adsabs.harvard.edu/abs/#1}{\textsf{ADS}}}\fi
\ifx \arxivurl  \undefined \def \arxivurl#1{\href{http://arxiv.org/abs/#1}{\textsf{arXiv}}}\fi
\ifx \urlurl  \undefined \def \urlurl#1{\href{http://#1}{\textsf{#1}}}\fi

\begin{document}

\begin{article}

\begin{opening}

\title{A Statistical Study of CME Properties and of the Correlation Between Flares and CMEs Over the Solar Cycles 23 and 24}

\author{A.~\surname{Compagnino}$^{1}$\sep
        P.~\surname{Romano}$^{2}$\sep
         F.~\surname{Zuccarello}$^{1}$\sep   
       }
\runningauthor{A.Compagnino \etal}
\runningtitle{Correlation Between Flares and CMEs Over the Solar Cycle 23 and 24}

   \institute{$^{1}$ Dipartimento di Fisica e Astronomia - Universit\'{a} di Catania, Via S. Sofia 78, 95123, Catania, Italy\\          
              $^{2}$ INAF - Osservatorio Astrofisico di Catania, Via S. Sofia 78, 95123, Catania, Italy              
              \href{mailto:email_address}{email: acomp@oact.inaf.it}
         }

\begin{abstract}
\label{S-Abstract}
We investigated some properties of coronal mass ejections (CMEs), such as speed, acceleration, polar angle, angular width and mass, using data acquired by the {\it  Large Angle Spectrometric coronagraph} (LASCO) 
onboard {\it Solar and Heliospheric Observatory} (SOHO)  from 31 July 1997 to 31 March 2014, \ie during  the 
Solar Cycles 23 and 24. We used two CME catalogs: one provided by the {\it Coordinated Data Analysis 
Workshops} (CDAW) Data Center and one obtained by the {\it Computer Aided CME Tracking software} (CACTus) 
detection algorithm. For each dataset, we found that the number of CMEs observed during the peak of Cycle 24 
was higher or comparable to the one during Cycle 23, although the photospheric activity during Cycle 24 was 
weaker than during Cycle 23.  Using the CMEs detected by CACTus we noted that the number of events 
$[N$] is of the same order of magnitude during the peaks of the two cycles, but the peak of the CME distribution 
during Cycle 24 is more extended in time ($N$ $>$ 1500 during 2012 and 2013). We ascribe the discrepancy 
between CDAW and CACTus results to the observer bias for CME definition in the CDAW 
catalog. 
   We also used a dataset containing 19,811 flares of C-, M-, and X-class, observed by the {\it Geostationary Operational Environmental Satellite} (GOES) during the same period. Using both datasets, we studied  the relationship between the mass ejected by the CMEs and the flux emitted during the corresponding flares:  we found 11,441 flares that were temporally correlated with CMEs for CDAW and 9120 for CACTus. Moreover, we found a log--linear relationship between the flux of the flares integrated from the start to end in the 0.1\,--\,0.8 nm range and the CME mass.
We also found some differences in the mean CMEs velocity and acceleration between the  events associated with flares and those that were not. 

\end{abstract}

\keywords{Coronal Mass Ejections, Initiation and Propagation; Flares, Dynamics.}
\end{opening}


\section{Introduction}
     \label{S-Introduction} 
		
Many studies of eruptive phenomena occurring in the solar atmosphere, such as flares, filament eruptions and coronal mass ejections (CMEs) are aimed at understanding the role  the global and local magnetic fields play in their triggering. According to most recent theoretical and observational works, flares, filaments, and CMEs are all manifestations of the same physical phenomenon:  magnetic reconnection. However,  the  temporal and spatial relationships among these events are still unclear. 
A preliminary aspect in this kind of investigation concerns the properties of CMEs.

For instance, \cite{Ivanov2001} studied the semiannual mean CME velocities for the time interval 1979\,--\,1989 and revealed a complex cyclic variation with a peak at the solar-cycle maximum and a secondary peak at the minimum of the cycle. The growth of the mean CME velocity is accompanied by a growth of the mean CME width. Moreover, they concluded that the secondary peak of the semiannual mean CME velocity in 1985\,--\,1986 is due to a significant contribution of fast CMEs with a width of about $100^{\circ}$ at the minimum of the cycle. This peak is supposed to be due to the increasing role of the global large-scale magnetic-field system at the minimum of the solar cycle.

\cite{Chen2006} reexamined whether flare-associated CMEs and filament eruption-associated CMEs have distinct velocity distributions and investigated which factors may affect the CME velocities. They divided the CME events observed from 2001\,--\,2003 into three types: the flare-associated type, the filament eruption associated type, and the intermediate type. For the filament eruption associated CMEs, the speeds were found to be strongly correlated with the average magnetic field in the filament channel.

\cite{Cremades2007} presented a survey of events observed during the period 1980\,--\,2005 and found that the latitude of a CME matches well with the location of coronal streamers, in agreement with \citet{hun93}.

More recently, \cite{Mittal2009a} have analyzed more than 12,900 CMEs observed by the { \it Solar and Heliospheric Observatory} (SOHO)/ { \it Large Angle Spectrometric coronagraph} (LASCO) during the period  1996\,--\,2007. They found that the speeds decrease in the decay phase of Solar Cycle 23. There is an unusual drop in speed in the year 2001 and an abnormal increase in speed in the year 2003. This increase corresponded to the so called Halloween events, \ie the high concentration of CMEs, X-class flares, solar energetic particle (SEP) events and interplanetary shocks observed during the months of October and November of that year. The same dataset showed that about 66\,\% of CMEs have negative acceleration, 25\,\% have positive acceleration and the remaining 9\,\% have very low acceleration \citep{Mittal2009b} in the outer corona.

Some difficulties in understanding  the relationships between flares and CMEs are due to  the different methods of observation that must be used to investigate these phenomena. In fact, the coronagraphs used to observe the outer corona where the CMEs are detected occult the solar disk and do not allow one to observe the source region where the corresponding flares take place. Therefore, many authors have tried to study these relationships from a statistical point of view.
     
For example, \cite{st.cyr91} considered, a dataset of two years, \ie from 1984 to 1986 acquired by the {\it Solar Maximum Mission} (SMM: \inlinecite{ph90}) (see Table \ref{T1}), found that 76\,\% of the CMEs were associated with erupting prominences, 26\,\% with H$\alpha$ flares, and 74\,\% with flares observed in the X-ray range. 
\cite{gil2000}, analyzing 18 CMEs observed by LASCO-C2 \citep{bec1995} and the ground-based {\it Mark-III K-Coronameter} (MK3) at the {\it Mauna Loa Solar Observatory} (MLSO: \citealp{mcf83}; \citealp{cyr99}) and 54 flares observed in H$\alpha$ during  two years of observation, between 1996 and 1998, found that 94\,\% of H$\alpha$ flares were associated  with CMEs and that 76\,\% of the CMEs were associated with eruptive prominences. Moreover, analyzing the same period, \cite{subd2001} using LASCO and the \textit{Extreme Ultraviolet Imaging Telescope} (EIT) instrument onboard \soho, \citep{del95})  found that 44\,\% of the CMEs were associated with eruptions of prominences embedded in an active region, while 15\,\% of those with eruptions occurred outside active regions. 

 \cite{zho2003} used data taken by LASCO onboard \soho from 1997 to 2001 and selected 197 front-side halo CMEs. They found that 88\,\% of those CMEs were associated with flares, while 94\,\% were associated with eruptive filaments. For 59\,\% of the CMEs, their initiation seemed to precede the associated flare onset recorded by GOES: \citealp{how74} ; \citealp{ludjon81}), while 41\,\% of the CMEs seemed to follow the flare onset.

\begin{table}
\caption{Previous results on the correlation between CMEs, flares and eruptive prominences.}
\label{T1}
\begin{tabular}{lccccc}     
Authors           & Number            & Period     & CMEs                 & CMEs             & CMEs         \\
                 & of                 &            & associated with             & associated with        & associated with  \\
								 & events             &            & eruptive  prominences          & H$\alpha$ flares           & X-ray  flares \\
\hline
\cite{st.cyr91} & 73 CMEs                 & 1984\,--\,1986  & 76\,\%                 & 26\,\%             & 74\,\% \\
\cite{gil2000}  & 18 CMEs    & 1996\,--\,1998  & 76\,\%                 & 94\,\%             &       \\
\cite{subd2001} & 32 CMEs                 & 1996\,--\,1998  & 59\,\%                 &                  &        \\
\cite{zho2003}  & 197 CMEs                & 1997\,--\,2001  & 94\,\%                 & 88\,\%             &        \\

  \hline
\end{tabular}
\end{table}

 Many authors (see, for example,\cite{gop2015} \cite{gop2015b}, \cite{you2012}) investigated  possible relationships between the CMEs physical parameters and the flare properties not excluding the narrow CMEs and using only the { \it Coordinated Data Analysis Workshops} (CDAW) dataset, even though the CACTus catalog was developed around 2004. The authors point out that the list is necessarily incomplete because of the nature of identification. In the absence of a perfect automatic CME-detection program, the manual identification is still the best way to identify CMEs \citep{you2012}.\\
Using the LASCO and EIT data taken by the \soho spacecraft, \cite{zha2001a} analyzed four events and found that the impulsive acceleration phase of the selected CMEs coincided well with the rise phase of the associated X-ray flares. Later, \cite{qiu2005} studied 11 events with varying magnetic-field configurations  in the source regions and concluded that the CMEs velocities were proportional to the total magnetic-reconnection flux, while their kinetic energy was probably independent of the magnetic configuration of the source regions.

An in-depth analysis of the correlation between X-ray flares and CMEs using GOES and LASCO archives from 1996 to 2006 has been performed by \cite{aar2011}. They considered 13,682 CMEs and selected 826 flare--CME pairs. They found that the CME mass increases with the flare flux, following an approximately log--log relationship: log(CME mass) $=$ 0.70 × log(flare flux), while the CME mass appears unrelated to their acceleration. 
 \cite{aar2011} also noted that CMEs associated with flares have higher average linear speeds ($495 \pm 8$ km\,s$^{-1}$) and negative average acceleration ($−1.8 \pm$  0.1 m\,s$^{-2}$), while CMEs not associated with flares have lower average linear speed ($422 \pm 3$ km\,s$^{-1}$) and  marginally positive average acceleration ($0.07 \pm$ 0.25 m\,s$^{-2}$). Finally, the width of CMEs resulted to be  directly correlated with the flare flux:  X-class flares are associated with the widest CMEs ($80^{\circ} \pm  10^{\circ}$),while B-class flares are  associated with the narrowest CMEs  ($42^{\circ}$ $\pm$  $1.4^{\circ} $).
	 
In this context, we intend to provide a further contribution to the knowledge of CMEs properties and of the correlation between flares and CMEs. In this article we present results obtained from the analysis of a dataset more extended than that of \cite{aar2011}, including 22,876 CMEs and 19,811 flares of GOES class C, M, and X observed from  31 July 1996 to  31 March 2014. We investigate how some previous mentioned relationships vary with the solar cycle.  In the next Section we describe our dataset, in Section 3 we show our results, and in Section 4 we discuss the results and draw our conclusions.

\urlstyle{same}


\section{Data Description}
      \label{S-data}  
			
The CMEs data used in this work were acquired by LASCO-C2 and -C3 coronagraphs onboard \soho during about   17 years (from  31 July 1997, to 31 March 2014). We used the CDAW Data Center online CME catalog, which is available at the following link: \href{http://cdaw.gsfc.nasa.gov/CME_list/}{\textsf{cdaw.gsfc.nasa.gov/CME\_list}} and the catalog obtained by the {\it Computer Aided CME Tracking software} (CACTus), which is available at the following link: 
 \href{http://sidc.oma.be/cactus/}{\textsf{sidc.oma.be/cactus/}}
 The use of both catalogs allowed us to compare the results from manual identification of CMEs (CDAW) and automatic tracking of CMEs (CACTus).
The CDAW catalog contains information on several CMEs parameters, such as their central polar angle (PA), \ie the mean angle between the two outer edges of the CME width measured counterclockwise from Solar North in degrees on the  plane of the sky, the linear velocity, the acceleration, the width, the mass, and the energy. The CACTus catalog contains all of information of CDAW catalog except the mass, the energy and the acceleration of the events. We remark that the CME start time reported in these catalogs corresponds to the first detection of the CME in the LASCO-C2 field of view, \ie the region from 2.0 to 6.0 solar radii. The whole CDAW dataset contains 22,876 events, while CACTus contains 15,515 events. 
We also considered 19,811 flares of C-, M-, and X-class that occurred during the same  time interval were observed by GOES. We used the reports of the National Geophysical Data Center (NGDC:   \href{ftp://ftp.ngdc.noaa.gov/}{\textsf{ftp://ftp.ngdc.noaa.gov/}})
to collect the main information, such as the time of the beginning, maximum, and end of each flare, the GOES X-ray class and the integrated flux from the beginning to the end. We considered 17,712 C-class flares, \textit{i.e.} 89.40\,\%, 1884 M-class flares, \ie 9.51\,\%, and 155 X-class flares, \ie 0.78\,\%. We also took into account the location on the solar disc where the flares occurred for 10,742 flares (about 50\,\%),  when this information was reported. 


\section{Results}
\subsection{CME Parameters Over the Solar Cycles}
\label{S-CME parameters over the solar cycles}

\begin{figure*}
\centering
\includegraphics[width=0.70\textwidth]{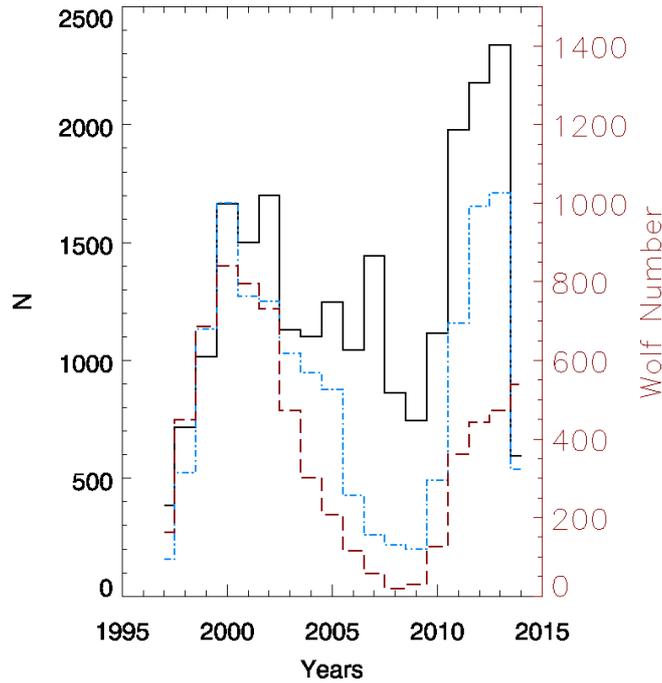}
\caption{Number of CMEs per year observed by LASCO during the selected time interval. The black and dot--dashed-blue lines indicate the results for the CDAW and CACTus catalogs, respectively.  The long-dashed-red line indicates the yearly Wolf Number during the same period}.  
\label{fig1}
\end{figure*}

In order to show the occurrence distribution of CMEs during the solar-activity cycle, we report in Figure \ref{fig1} the number of CMEs [$N$] observed by LASCO during each year of our observation time interval (the black and the dot--dashed blue lines correspond to the data obtained from the CDAW and CACTus catalogs, respectively). We clearly see two peaks corresponding to the maxima of the Solar Cycles 23 and 24. In particular, we note that for the CDAW dataset the peak of  Cycle 24 is higher than the one referring to Cycle 23.  This result seems to be in contrast with the fact that the magnetic activity during  Cycle 24 was weaker than during Cycle 23 \citep{tjh2015}, as shown in Figure \ref{fig1} (long-dashed-red line), where the yearly Wolf number is reported.
Moreover, using the CMEs detected by CACTus we note that N is of the same order of magnitude during the peaks of the two cycles, and that the peak of CME distribution during the Cycle 24 is more extended in time ( $N$ $>$ 1500 during 2012 and 2013). The different results obtained from CDAW and CACTus could be ascribed  to the observer bias in the CME definition in the CDAW catalog (\citealp{rob09}; \citealp{weho2012}; \citealp{yas2008}).  Moreover as far as the higher or comparable number of CMEs recorded during Cycle 24, despite the low number of sunspots observed during this cycle, it is worth of note to recall that \cite{gop2015b} found a similar result from the analysis of halo CMEs.
  
\begin{figure*}
\centering
\includegraphics[width=0.55\textwidth]{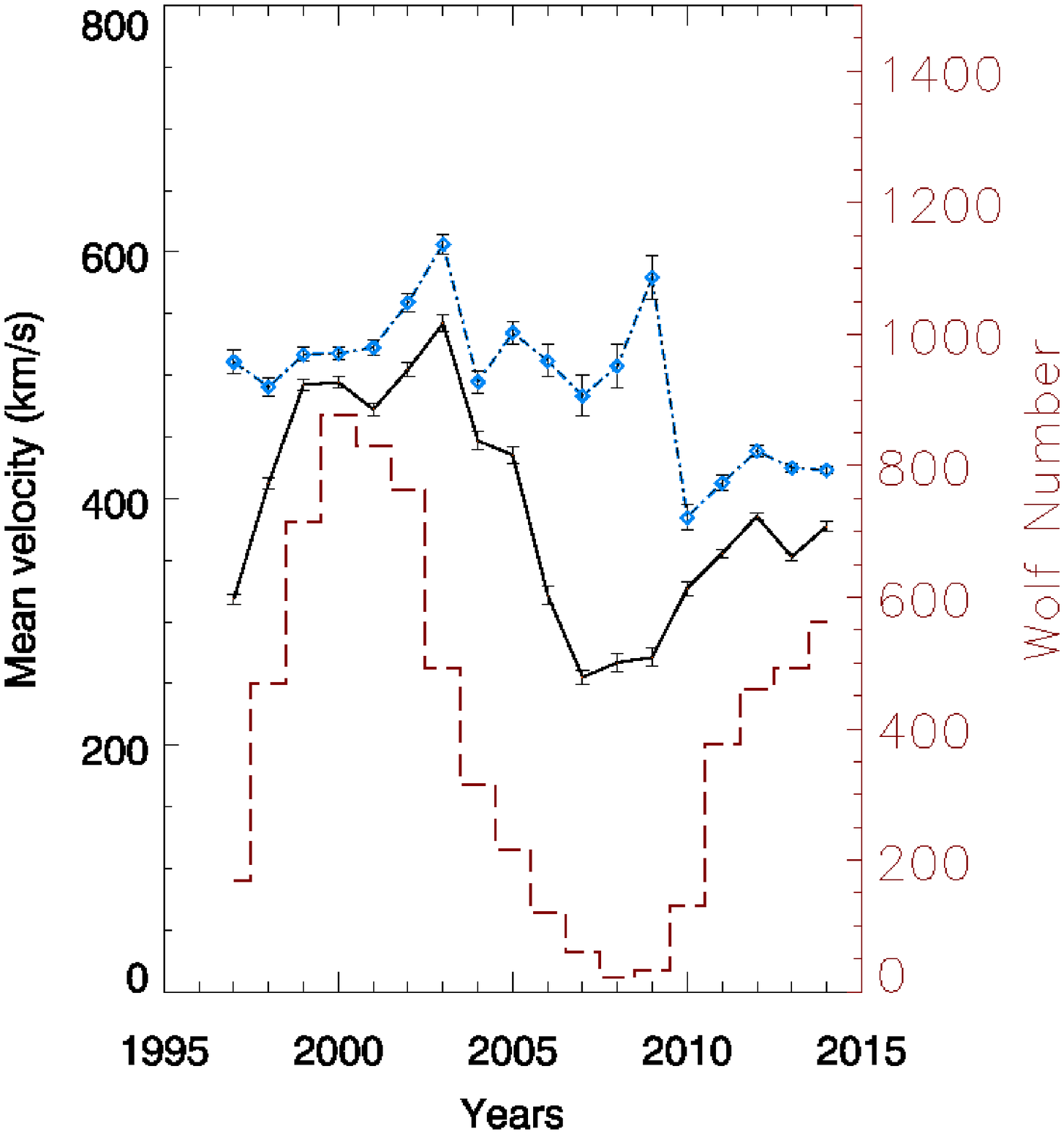}
\includegraphics[width=0.55\textwidth]{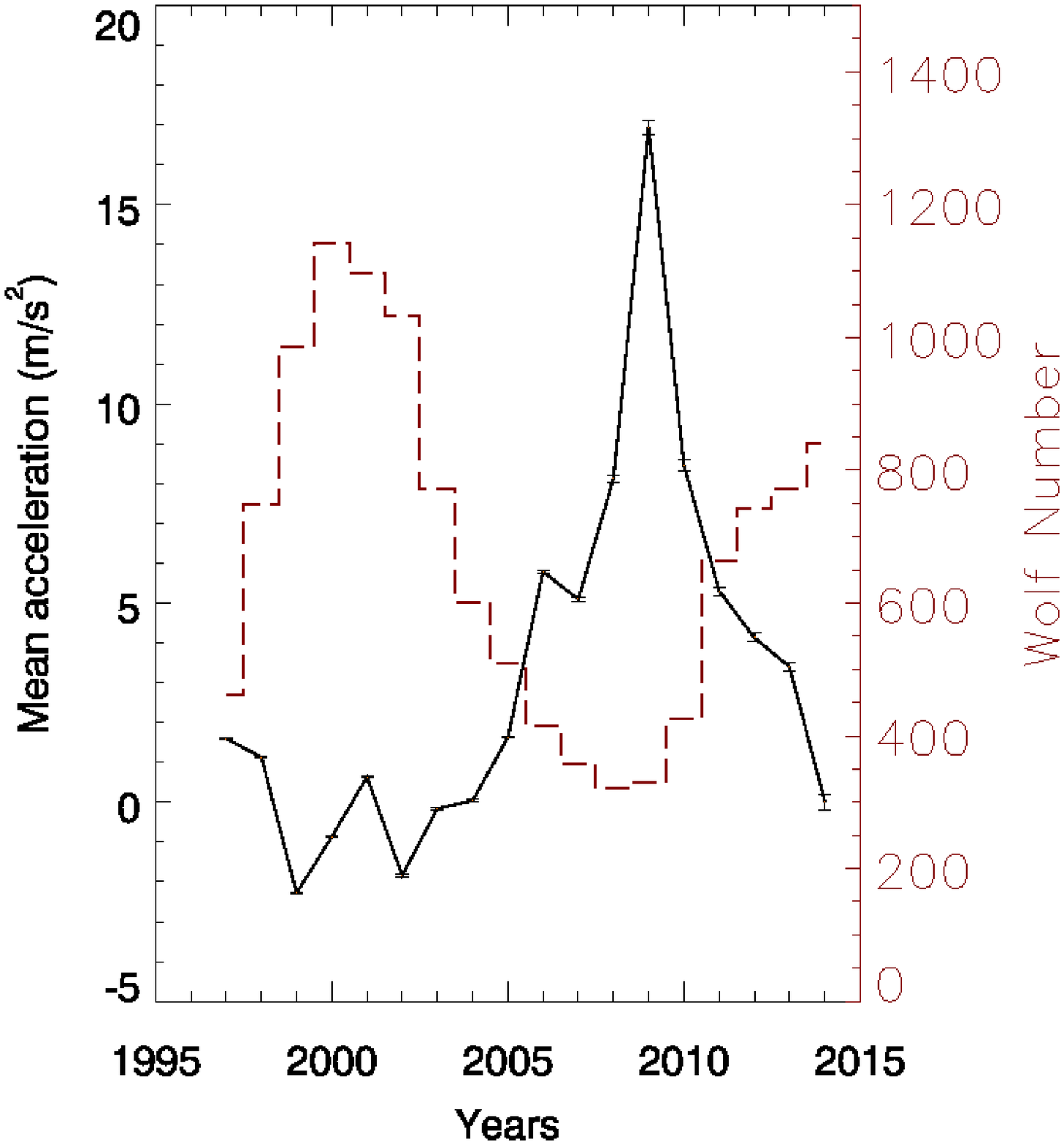}
\includegraphics[width=0.55\textwidth]{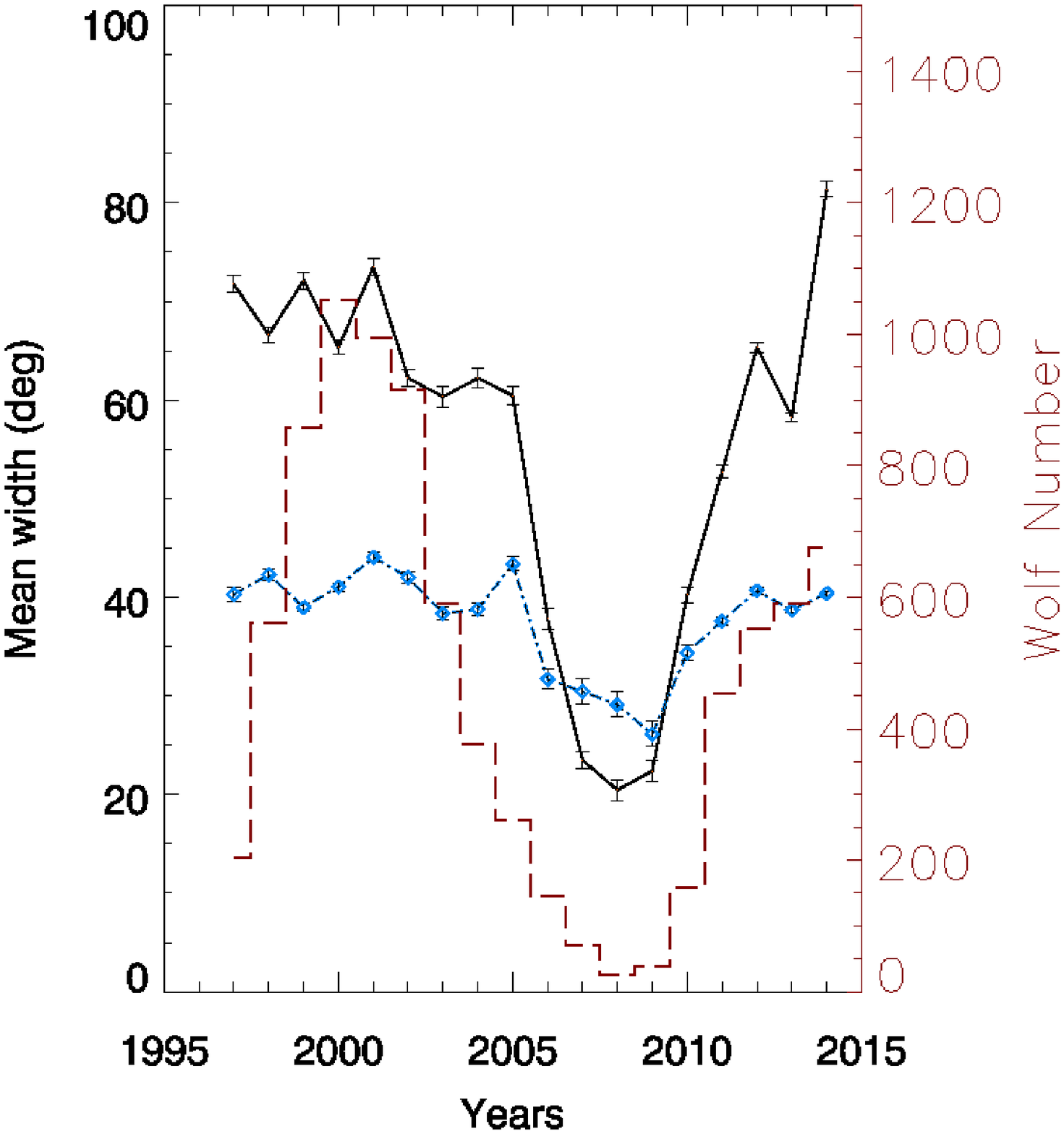}
\caption{Distribution of the mean annual CME velocity  (top panel),  acceleration (middle panel,) and angular width (bottom panel)  from 1996 to 2014. The black and dot--dashed-blue lines indicate the results for the CDAW and CACTus catalogs, respectively.  In each panel the yearly Wolf number is shown (long-dashed-red line). The blue diamonds indicate the CACTus data.} 
\label{fig2}
\end{figure*}

In Figure \ref{fig2} (from top to bottom) we show the behavior of the average annual velocity, acceleration and angular width for all the observed CMEs during the Solar Cycles 23 and 24. For the velocity and the angular width we plot the results of CDAW dataset (black line) and of CACTus dataset (dot--dashed-blue line), while for the acceleration we show only the CDAW results because this parameter is not available in the CACTus dataset. The blue diamonds indicate the CACTus data.
 
We calculate the CME annual mean velocity by considering the total number of CMEs for each year and then considering the velocity averaged with respect to this number. The same procedure was applied for the other parameters. We also computed the error bars for the mean distribution of these quantities, as $\sigma/ \sqrt{N}$, where $\sigma$ is the standard deviation of the corresponding quantity and N is the number of CMEs that occurred in each year.  The error bars are smaller than the symbol sizes, although the error for the acceleration is dominated by the uncertainty of the individual measurements. The red line in each panel shows the yearly Wolf number. For the average velocity we can see a first peak around 2003 and a second one between 2012 and 2014 in both of the datasets. The behavior of both of the datasets reflects the solar-activity cycles and can be interpreted according to \cite{qiu2005} as an effect of the magnetic flux involved by the events during the solar maxima. However, one difference between the two datasets is that in the mean-velocity distribution of the CMEs detected by CACTus there is another peak between 2008 and 2010, during the minimum of the solar cycle. A secondary peak was reported by \cite{Ivanov2001} from the analysis of CMEs velocities for the time interval 1979\,--\,1989,and was interpreted by these authors as due to a significant contribution of fast CMEs ($V>400$ km\,s$^{-1}$) with a width of $100^{\circ}$. \\
The mean CMEs acceleration for CDAW is $3.17 \pm$ 0.3 m\,s$^{-2}$.    
In the distribution of the CME average acceleration (middle panel) of Figure \ref{fig2}) we see negative values around the maximum of Solar Cycle 23 and a peculiar peak of about $15  \pm$ 2.71 m\,s$^{-2}$ in 2009. We note that this peak occurs approximately when we observe the minimum in the annual average velocity distribution for CDAW and the second peak of the velocity distribution for CACTus. However, we argue that statistically the slower CMEs are characterized by higher positive values of acceleration.\\ In Figure \ref{fig2} ( bottom panel), showing the mean angular width, the values in the years 2000\,--\,2003 for CDAW dataset ($\approx 20^{\circ}$) suggest that on average the narrowest CMEs are the slowest ones (compare with  Figure \ref{fig2} (top panel)) \citep{yas2008}. We note also that for the CACTus dataset the fastest CMEs are the narrowest \citep{yas2008}. Both datasets show a minimum in the mean width distribution over the years corresponding to the minimum of the solar cycle, although \cite{yas2008} found that the CACTus catalog has a larger number of narrow CMEs than CDAW. This difference in the two samples determines a different amplitude in the range spanned by the mean angular width distribution, \ie the CACTus catalog the mean width varies from $\approx 30^{\circ}$ during the minimum of solar-activity to $\approx 40^{\circ}$ during the maximum of activity, while for CDAW catalog  varies between  $\approx 20^{\circ}$ and $\approx 80^{\circ}$.
\begin{figure*}
\centering
\includegraphics[width=1.0\textwidth]{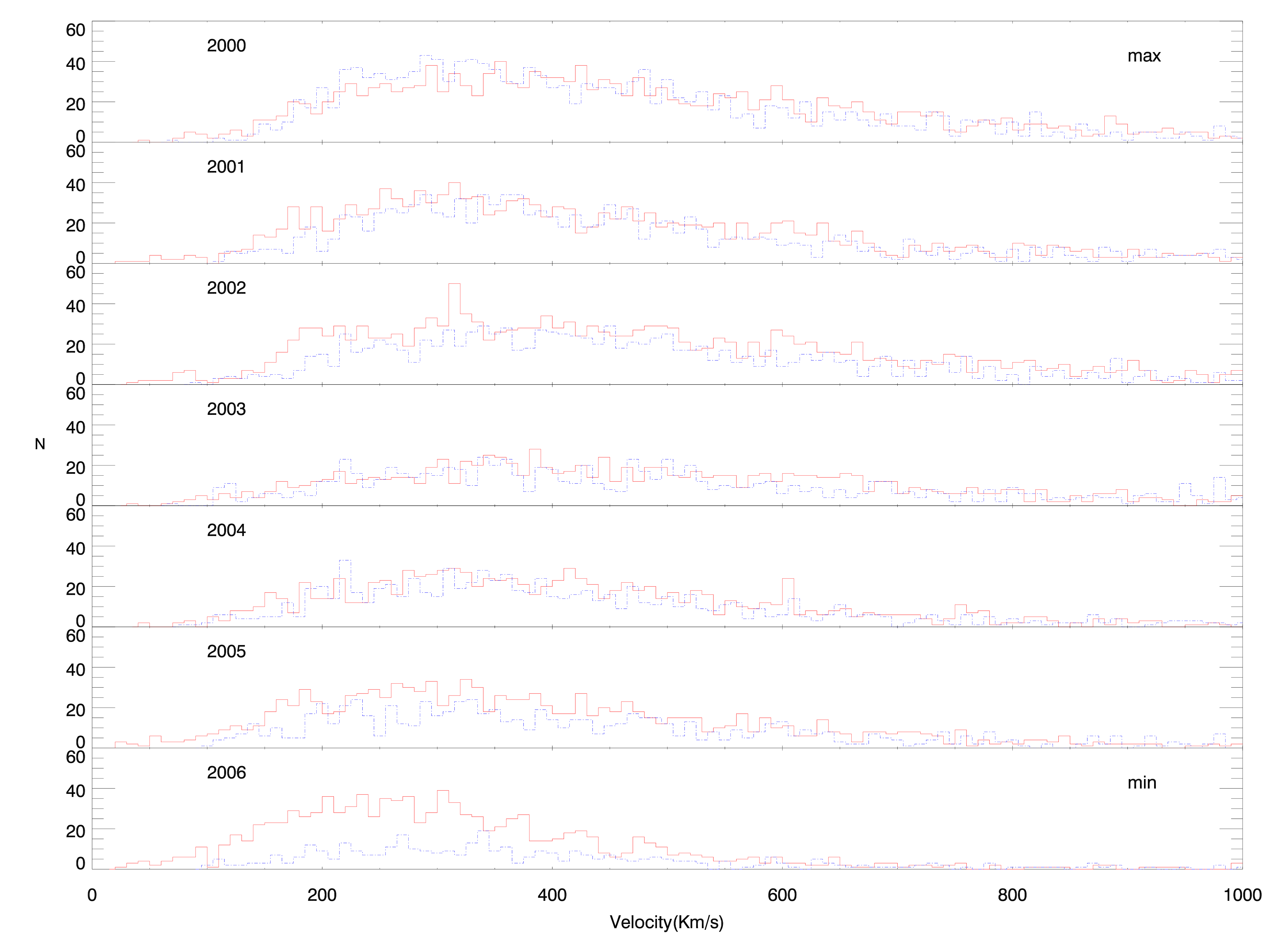}
\caption{Distribution of the velocity of the CMEs  in each year between 2000 and 2006  for CDAW (red line) and CACTus catalogs (dot--dashed-blue line).}
\label{fig3}
\end{figure*}

\begin{figure*}
\centering
\includegraphics[width=1.0\textwidth]{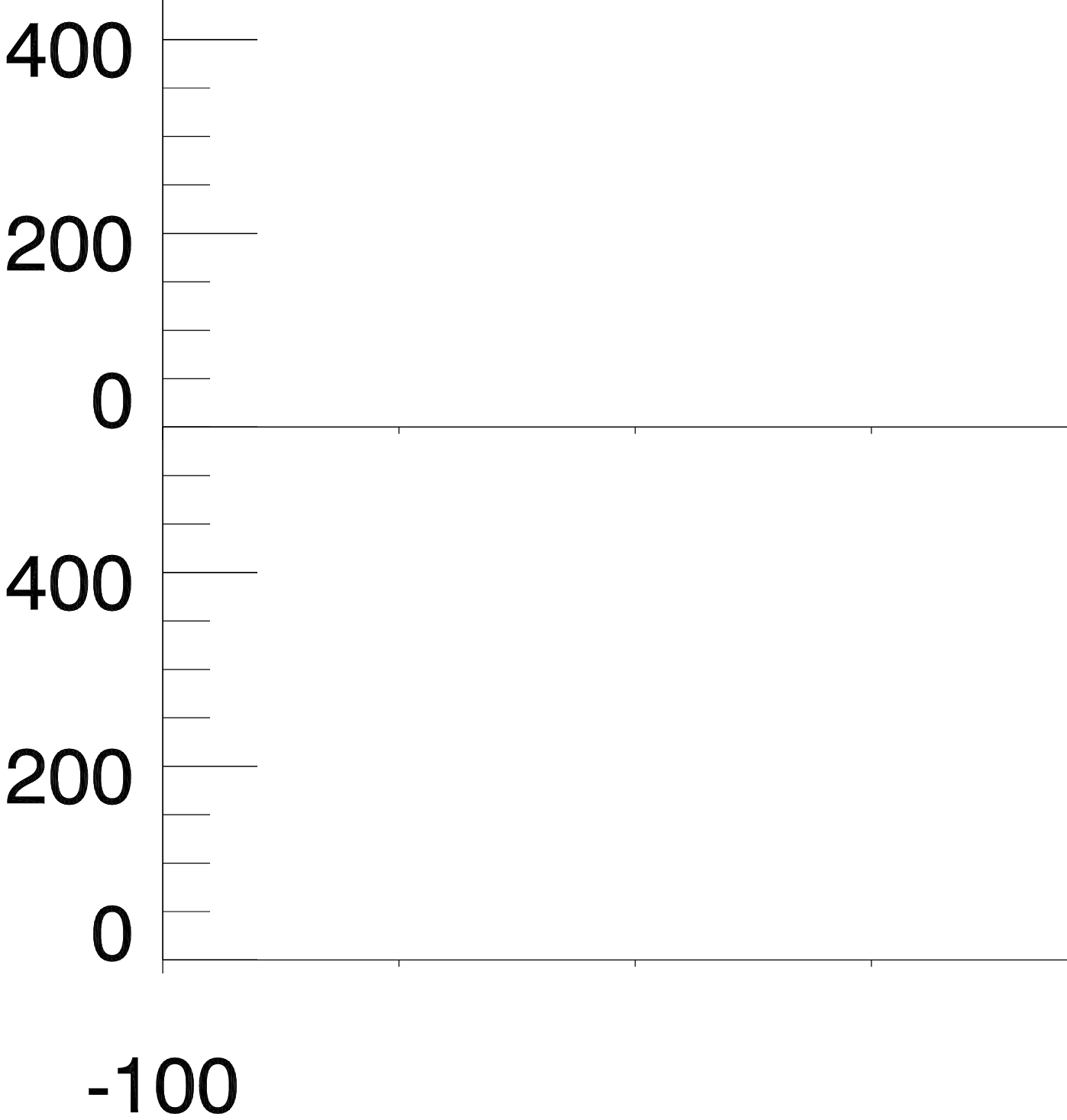}
\caption{Distribution of the acceleration of the CMEs in each year between 2000 and 2006.}
\label{fig4}
\end{figure*} 

\begin{figure*}
\centering
\includegraphics[width=1.0\textwidth]{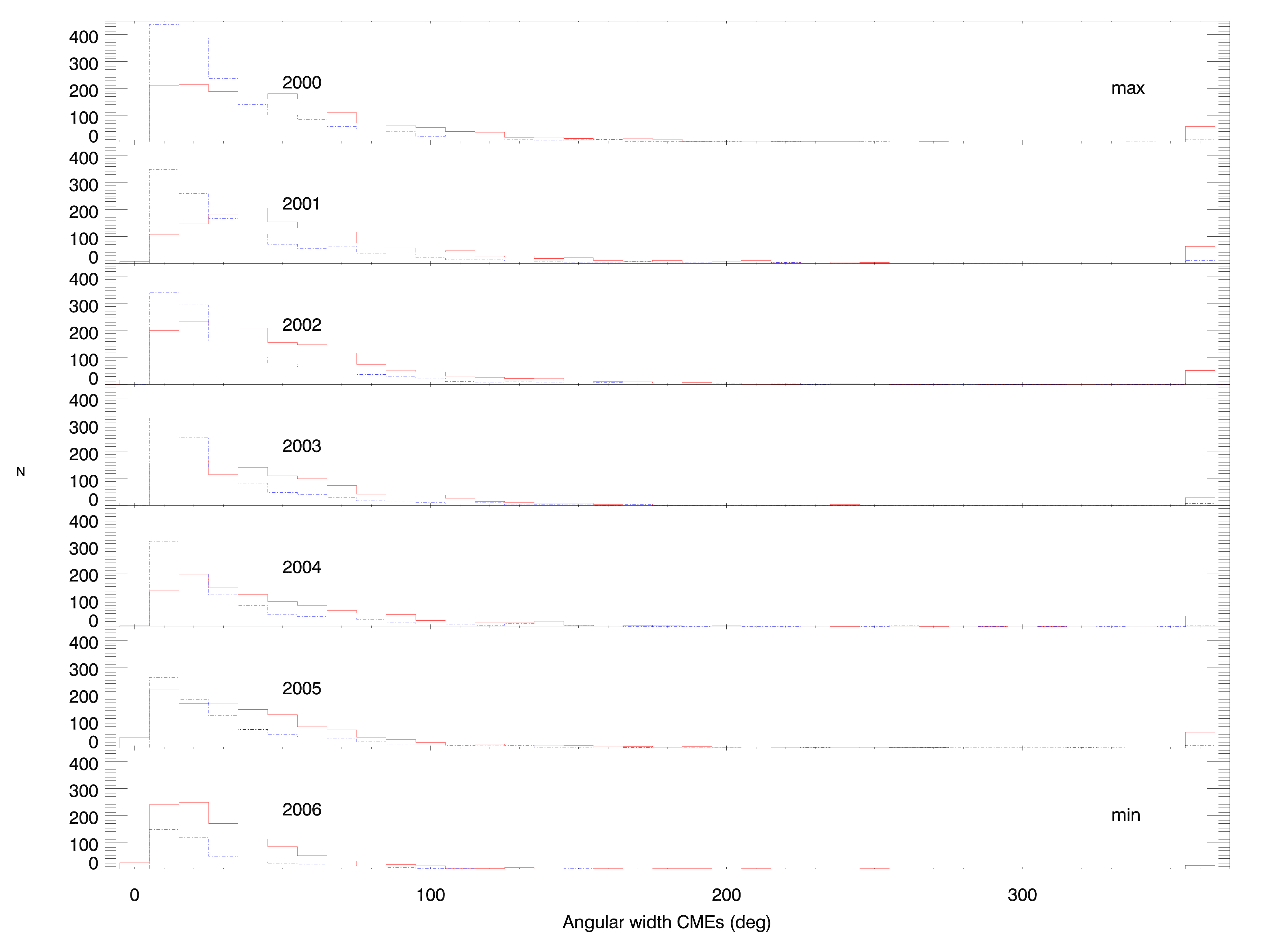}
\caption{Distribution of the CME angular width in each year between 2000 and 2006  for CDAW (red line) and CACTus (dot--dashed-blue line).  The latter bin refers to halo CMEs.}
\label{fig5}
\end{figure*}

\begin{figure*}
\centering
\includegraphics[width=1.0\textwidth]{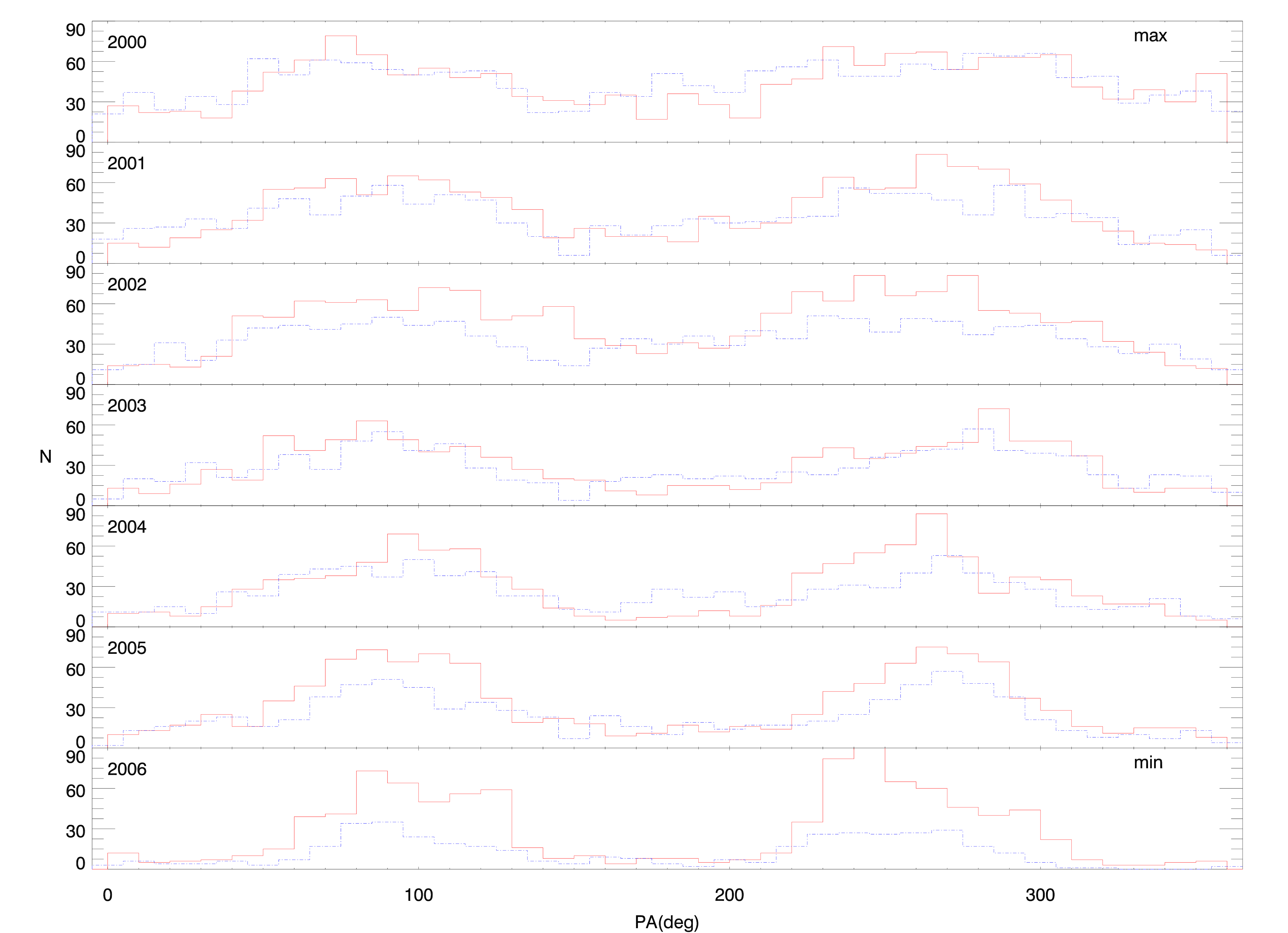}
\caption{Distribution of the polar angle (PA) along which the CMEs propagate in each year between 2000 and 2006 for CDAW (red line) and CACTus (dot--dashed-blue line). The PA is measured in degrees from the Solar North in counter-clockwise direction.}
\label{fig6}
\end{figure*}

In order to further analyze the properties of CMEs during the descending part of the Solar Cycle 23 (from 2000 to 2006), in Figures \ref{fig3},  \ref{fig4},  and \ref{fig5}  we show the distribution of the velocity, acceleration and width of the CMEs. 

Figure \ref{fig3} shows that the tail of the velocity distribution decreases from the maximum (2000) to the minimum (2006) of the cycle. This means that most of the fastest CMEs velocities ($>$ 1000 km\,s$^{-1}$) occur during the maximum of the solar cycle, when the amount of magnetic field  emerged in the atmosphere reaches higher values. The average velocity of the CMEs for CDAW during these years is 396.25 $\pm 1.67$ km\,s$^{-1}$, where this uncertainty is the standard deviation of the mean, while the mean-velocity for CACTus is 491.78 $\pm 2.66$ km\,s$^{-1}$. We also remark that in Fig \ref{fig3}  shows only a few CMEs characterized by velocities $\geq$ 600 km\,s$^{-1}$ (note that in these plots the velocities are not averaged on each year as in Figure \ref{fig2} (top panel)).

The CME accelerations in each year (Figure \ref{fig4}) are distributed mainly between -50 m\,s$^{-2}$ and 50 m\,s$^{-2}$ , and few events are characterized by higher values during the solar-cycle maximum. We note an increase of the difference between the negative and the positive accelerations while the solar activity decreases. Therefore, the average acceleration per year is mainly negative during the years of  higher activity and is positive during the minimum of the solar cycle, as already shown in Figure \ref{fig2} (middle panel). 

The angular-width distribution (see Figure \ref{fig5}) shows that the majority of the CMEs are characterized by an angular width lower than $100^{\circ}.$ We note that the CACTus dataset presents a greater number of CMEs narrower than $40^{\circ}$ and a smaller number of CMEs wider than $40^{\circ}$ in comparison with the CDAW dataset (see \citep{yas2008}).
The mean angular sizes (latitudinal extents) projected against the plane of the sky  of all CMEs of the CDAW datset is $55.78^{\circ} \pm 0.44 ^{\circ} $, \ie slightly greater than $50^{\circ}$ as found by \cite{cane2000}. The mean CMEs width for CACTus dataset is $39.48 ^{\circ} \pm 0.38 ^{\circ}$.
Only during the solar-cycle maximum (2000 and 2001) we observe a significant number of CMEs with a width larger than $100^{\circ}.$ The last bin of the plots in Figure~\ref{fig5} represents the halo CMEs contained in the CDAW dataset, \ie CMEs with angular width of $360^{\circ}$. Our dataset, composed by 22,876 CMEs, contains 616 halo CMEs, \ie 2.69\,\% of the total number of CMEs.

We used the same sample of years to study the variation in time of the distribution of the polar angle (PA) along which the CME propagated (Figure \ref{fig6}). For both datasets we can see two peaks centered around the PA corresponding to low latitudes, \ie where the ARs form and where the CMEs start. We note that the distribution of PA changes in time from a broader distribution in 2000 (near the maximum of Solar Cycle 23) to a more peaked distribution in 2006 (near the minimum of the solar-activity). 
We deduce that the latitude distribution of CMEs follows the latitude distribution of the closed-magnetic -field regions in the corona, which is consistent with the fact that the CMEs originate in closed-field regions \citep{hun93}.
	

\subsection{Flare Distribution Over the Solar Cycles}
\label{Flare distribution over the solar cycles}

\begin{figure*}[h!]
\centering
\includegraphics[width=0.70\textwidth]{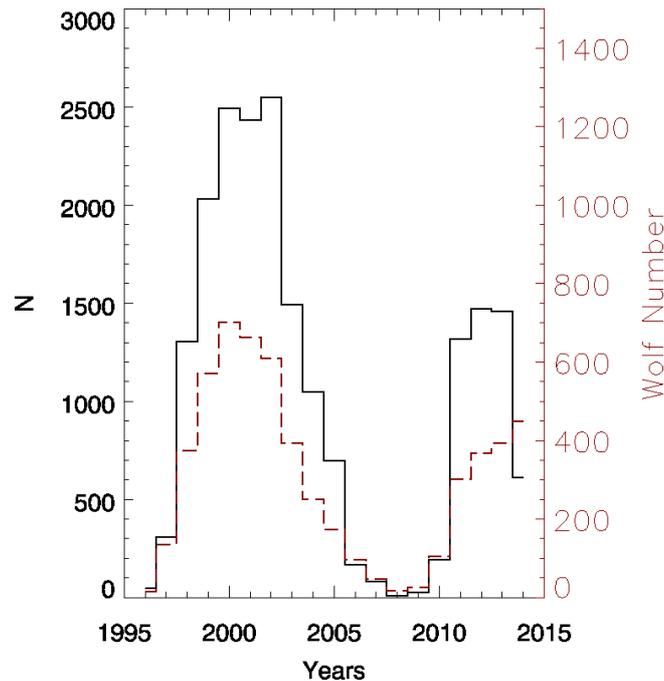}
\caption{Total-flares distribution in the selected time interval (black line) and yearly Wolf Number (long-dashed-red line).}
\label{fig7}
\end{figure*}

Using the dataset relevant to flares of C-, M-, and X-class observed by GOES, during the period 1996\,--\,2014, we determine the total number of flares per year. The result is shown in Figure \ref{fig7} (black line), where we can see two peaks approximately corresponding to the maxima of the solar-activity cycles. The maximum of Solar Cycle 23 is quite long: in 2000, 2001, and 2003 we observed about 2500 flares per year. We see also that the maximum of flare activity in 2014 corresponds to  the  peak of the Wolf Number in Cycle 24. In 2008 we observed only 11 flares: 10 of C class, and 1 of M class.

\begin{figure*}[h!]
\centering
\includegraphics[width=0.70\textwidth]{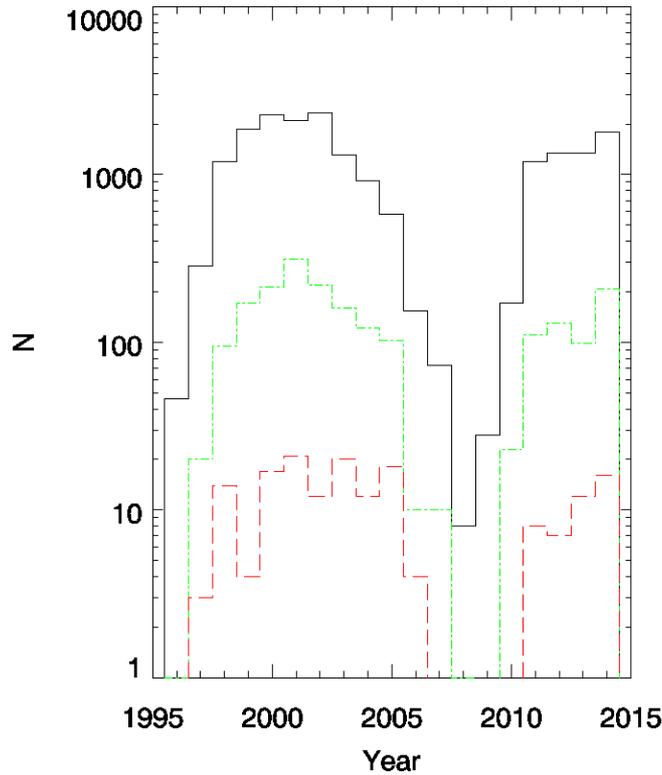}
  \caption{Distribution of C- (black), M- (dot--dashed-green line), and X-class (long-dashed-red line) flares in the selected time interval.}
\label{fig8}
\end{figure*}

When we distinguish among the C-, M-, and X-class flares (Figure \ref{fig8}) we see that the maxima of the M-class flares occurred in 2002 and in 2014 and that both the C- and the M-class flare distribution have a higher maximum during Cycle 23 than during  Cycle 24 (we recall that a different result was found for CMEs, see Figure \ref{fig1}). 


 \subsection{Correlation between flares and CMEs}
\label{Correlation between flares and CMEs}

\begin{figure*}
\centering
\includegraphics[width=0.65\textwidth]{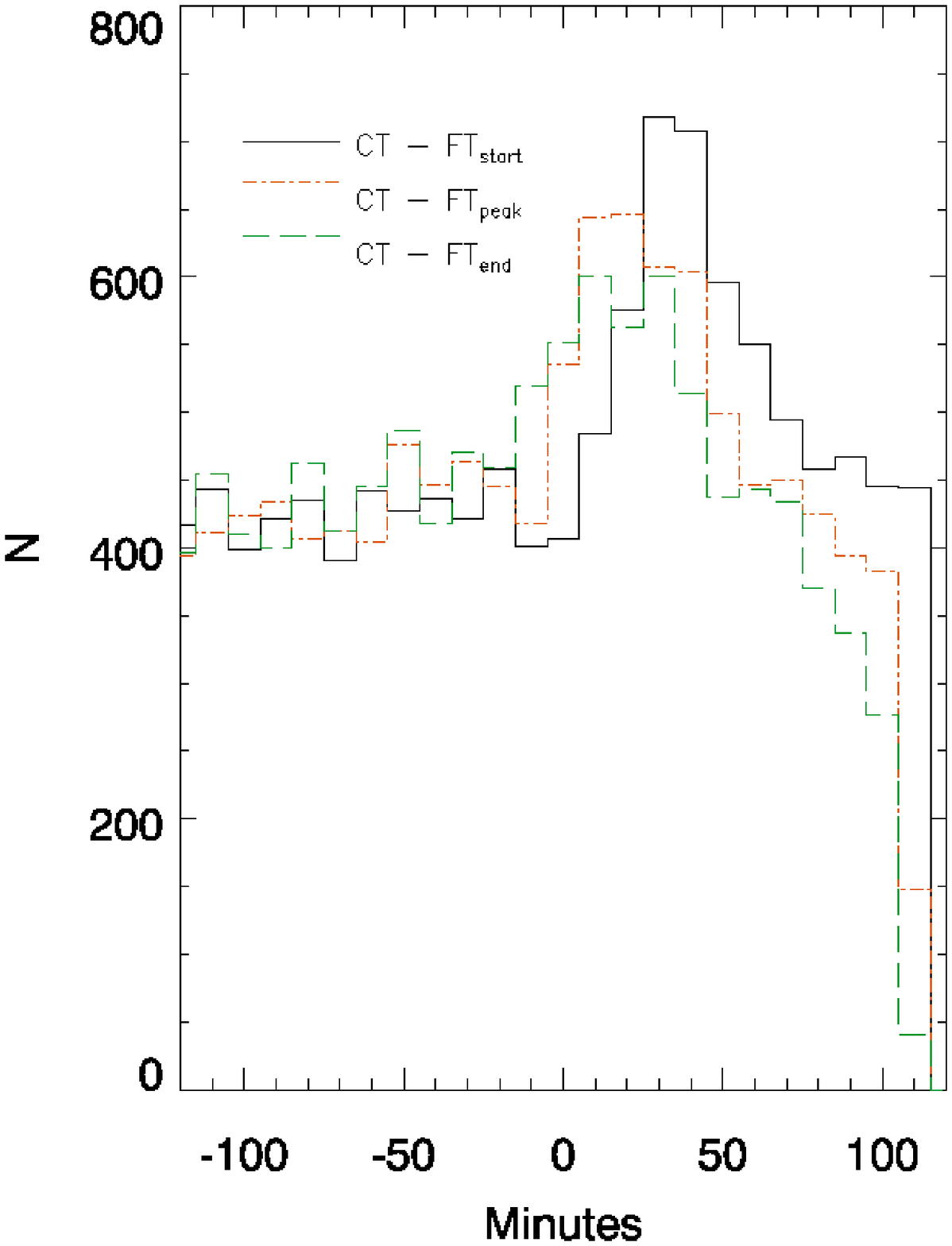}
\includegraphics[width=0.65\textwidth]{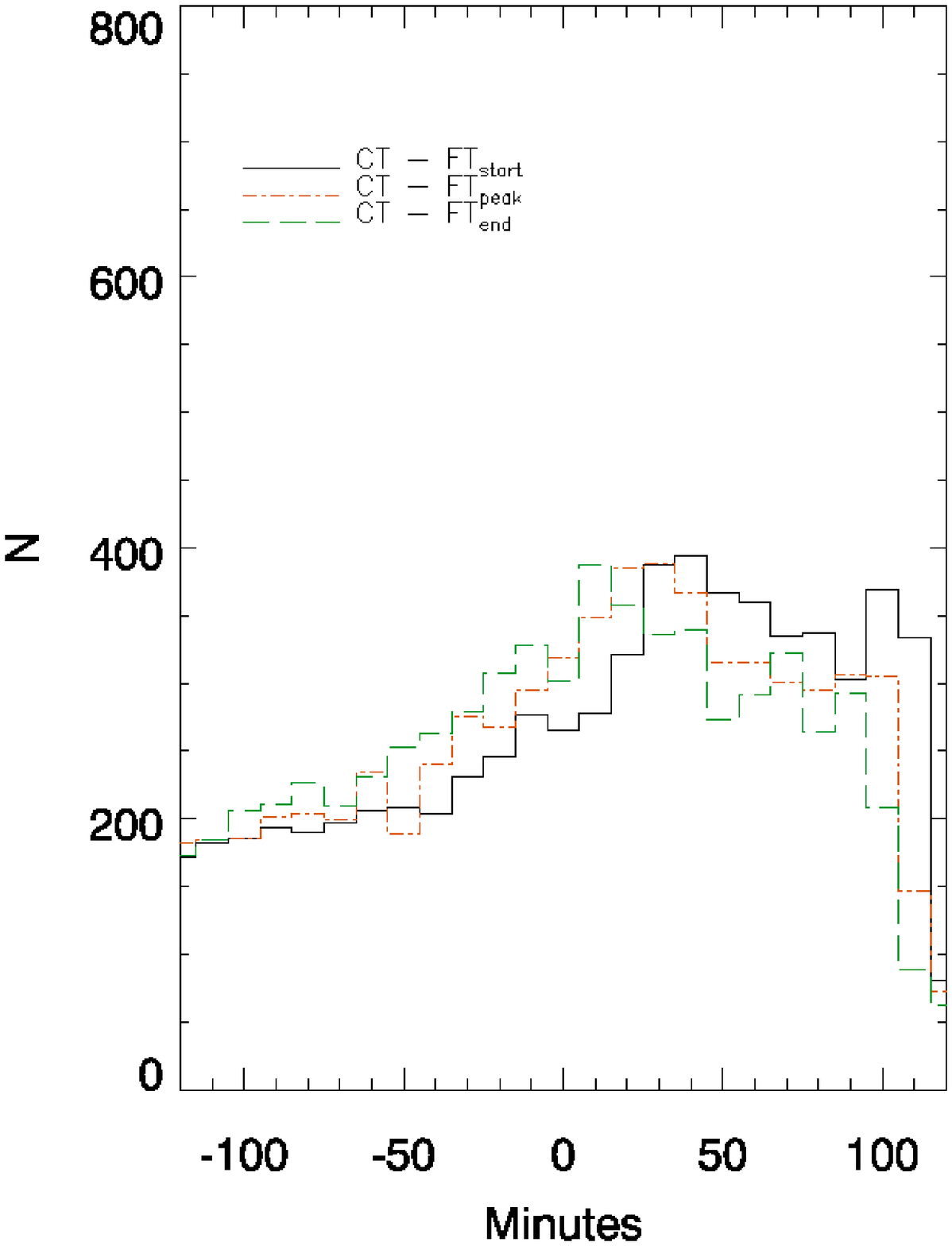}
\caption{ Distribution of the  CME-flare associated events as a function of the time difference between the CME first observation and the flare start (black line), peak (dot--dashed-red line), and end time (long-dashed-green line). CT and FT refer to CME first observation time and flare time, respectively. The  top and bottom panels refer to the CDAW and CACTus datasets, respectively. }
\label{fig9}
\end{figure*}

In order to determine the possible association between one CME and one flare, we used a temporal criterion, requiring that both flare and CME occur within a set time window. We initially set time windows to select CME first observation time that occurs within $\pm$ two hours of the flare start time, peak time, or end time, using both datasets. When we consider the CDAW dataset and the flare start time, we find that the highest number of CMEs and flares (59.57\,\%) is characterized by a difference in time between 10 and 80 minutes (see the black line in the  (top panel) of Figure \ref{fig9}), confirming the results obtained by \cite{aar2011}. The distribution of the CME-flare associated events for the CACTus dataset (bottom panel) of Figure \ref{fig9}) is slightly different from the CDAW dataset. In fact, we observe a wider time range (from -30 to 110 minutes) for the occurrence of most of the associated events. Considering the flare start time and a time interval of $\pm $ two hours, we found 11,441 and 9120 flares associated with CMEs using the CDAW and the CACTus catalogs, respectively. The distributions of these flares according to the GOES class are reported in Table \ref{T2}. In both cases the temporal shift between the flare and the CME decreases if we consider the distributions obtained using the flare-peak time and the flare-end time (see Figure \ref{fig9}).
In Table \ref{T2}, \ref{T3}, and \ref{T4} we present the associated CMEs-flares using only temporal criteria. In this way we obtained in first approximation a high number of CMEs associated with flares, which   drastically decreases when we apply a spatial correlation between these events to select the true-associated events as described in Section. \ref{S-CME parameters and flare energy}, so that finally we found only 1277 CMEs associated with flares that are spatially and temporally correlated.

\begin{table}
\caption{Correlation between CMEs and flares  for CDAW and CACTus datasets in the $\pm$ two hours time interval.}
\label{T2}

\begin{tabular}{cccccc}  
\multicolumn{2}{c}{Flares}   &   \multicolumn{2}{c}{CDAW} &  \multicolumn{2}{c}{CACTus}\\ 
\hline                   
GOES  & Number  & flare                & CMEs           & flare                &CMEs                   \\
class &   of    & associated           & not associated & associated           & not associated         \\
      & events  & with CME             & with flares    & with CME             & with flare [\%] \\                      
      &         & $\,\pm$ 2\,h [\%]&[\%]    & $\, \pm$ 2\,h [\%]&  \\             

\hline

C  & 17,712     & 10,003 (56.47\,\%)         & 11,074 (48.40\,\%)    & 7755 (43.78\,\%) & 6396 (41.22\,\%) \\ 
M  & 1884     & 1308 (69.43\,\%)   &                            & 1242 (65.92\,\%)                                 \\
X & 155       & 130 (89.39\,\%)    &                            & 121 (78.06\,\%)                                  &        \\
Halo CMEs  & 616                 & 315 (51.14\,\%)              & 301(48.86\,\%)  & &      \\

\\

  \hline
\end{tabular}

\end{table}

The same analysis was performed using the time intervals $\pm $ one hour, $\pm 30 $ minutes, and the flare start time. The results are reported in Tables \ref{T3} and \ref{T4}. 
Similar behavior of the correlation between CMEs and flares in different time intervals and in different GOES classes has been found for both datasets. We ascribe the smaller number of CACTus CMEs associated with X-class flares to the limits of the CACTus algorithm in the detection of the fastest CMEs \citep{yas2008}.

\begin{table}
\caption{Correlation between CMEs and flares  for CDAW and CACTus datasets in the $\pm$ one-hour time interval.}
\label{T3}

\begin{tabular}{cccc}     
   
  \multicolumn{2}{c}{Flares}  & \textbf{CDAW} & \textbf{CACTus}\\
\hline
GOES  & Number  & flare                 & flare                 \\
class &   of    & associated            & associated            \\
      & events  & with CME              & with CME              \\                      
      &         & $\pm$ 1\,h [\%] & $\pm$ 1\,h [\%] \\

\hline

C & 17,712                  & 5842 (32.98\,\%) &  4228 (23.87\,\%)                        \\
M  & 1884     & 951 (50.48\,\%)                &   771 (40.92\,\%)    \\
X & 155                 & 118 (76.13\,\%)       &    86 (55.48\,\%)                            \\
\hline
\end{tabular}
\end{table}

\begin{table}
\caption{Correlation between CMEs and flares \textbf{for CDAW and CACTus datasets} in the $\pm$ 30\,min. time interval.}
\label{T4}
\begin{tabular}{lccc}     

 \multicolumn{2}{c}{Flares}  & \textbf{CDAW} & \textbf{CACTus}\\                   
\hline

GOES  & Number  & flare                 & flare                 \\
class &   of    & associated            & associated            \\
      & events  & with CME              & with CME              \\                      
      &         & $\pm$ 30\,min. [\%] & $\pm$ 30\,min. [\%] \\

\hline

C & 17,712                  & 2992 (16.89\,\%) &   2159 (12.19\,\%)                       \\
M  & 1884     & 445 (23.62\,\%)                & 341 (18.099\,\%)     \\
X & 155                 & 62  (40.00\,\%)       &  37  (23.87\,\%)                             \\
\hline
\end{tabular}
\end{table}

\begin{figure*}
\centering
\includegraphics[width=0.44\textwidth]{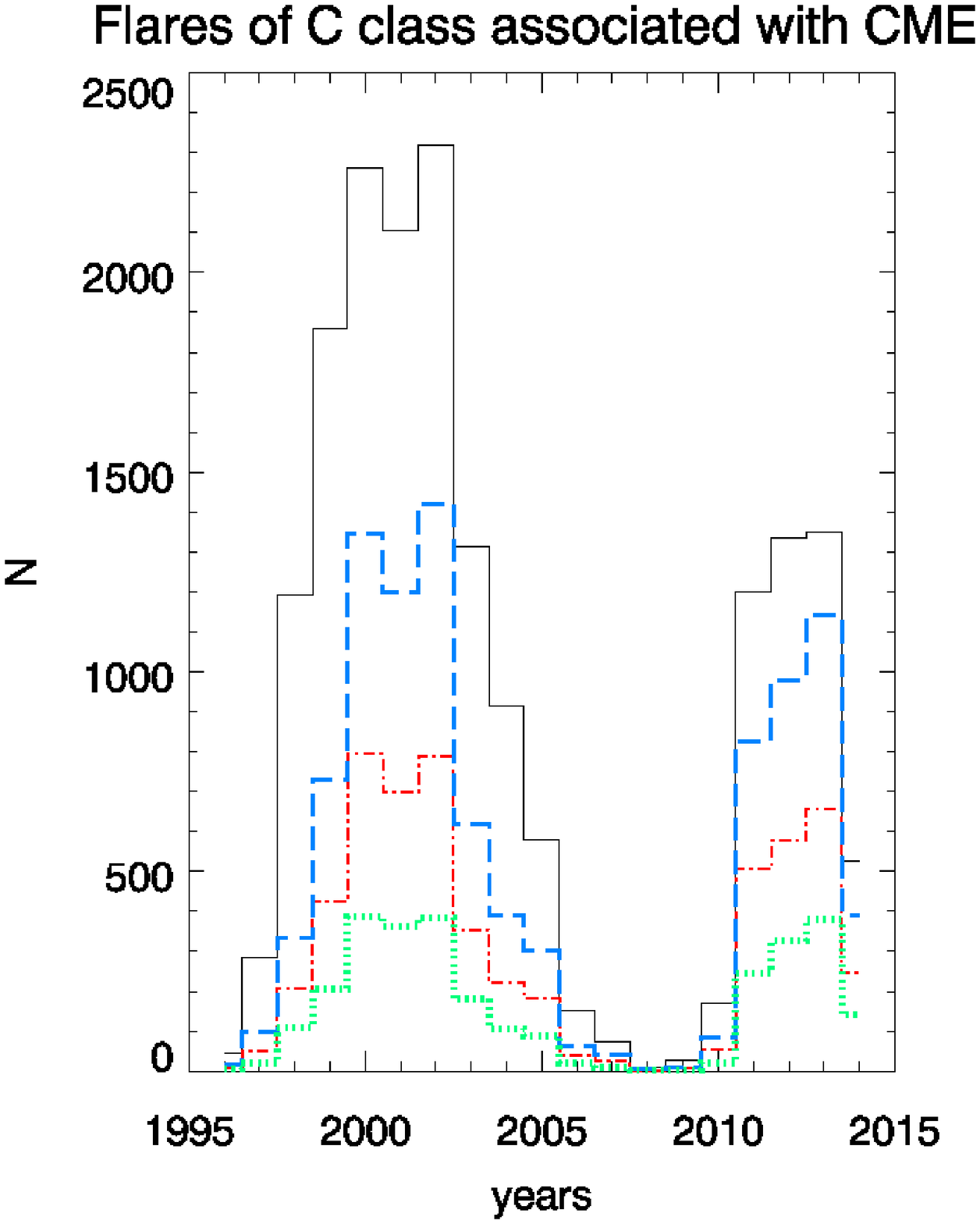} \\
\includegraphics[width=0.44\textwidth]{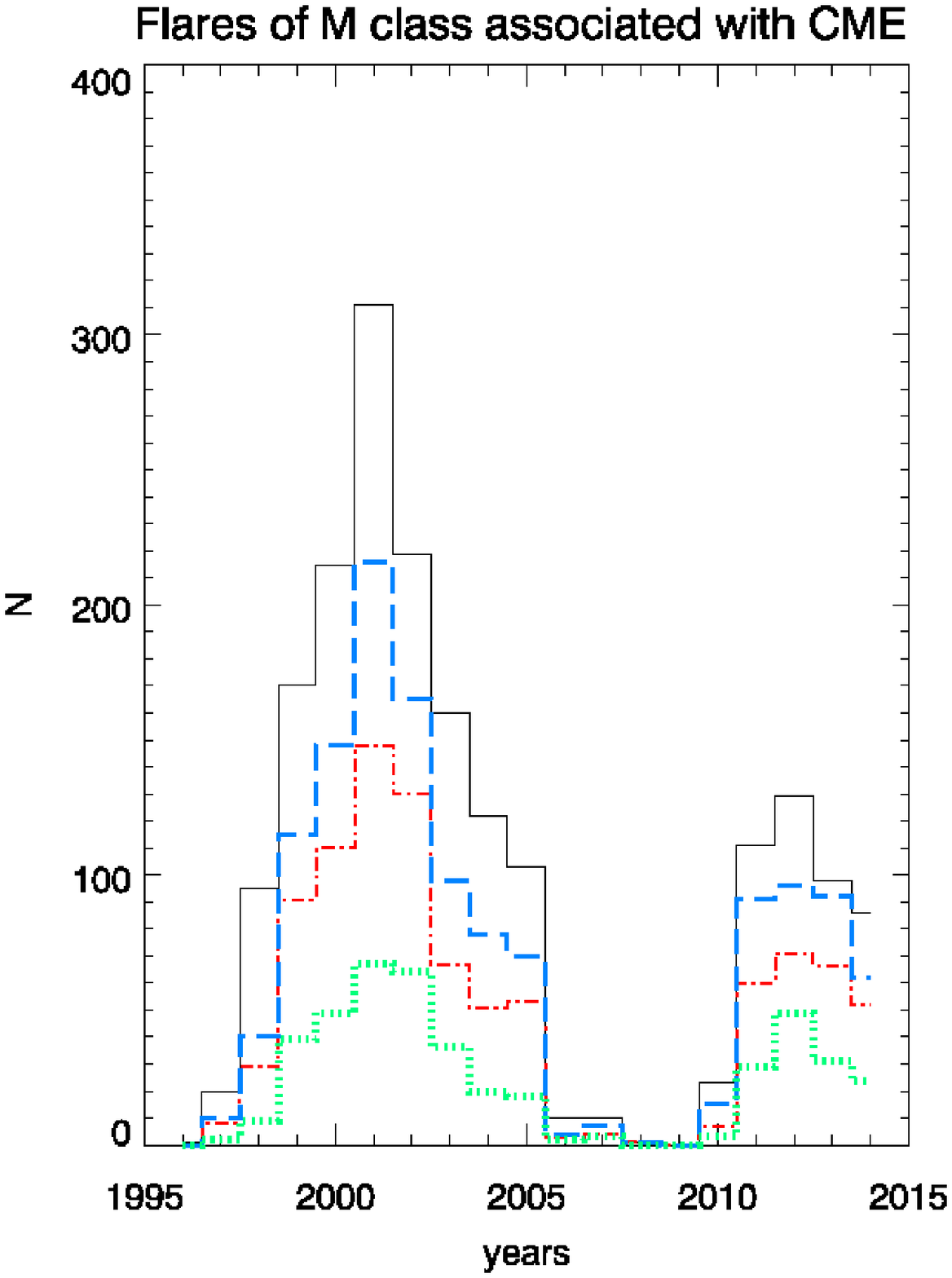}\\
\includegraphics[width=0.44\textwidth]{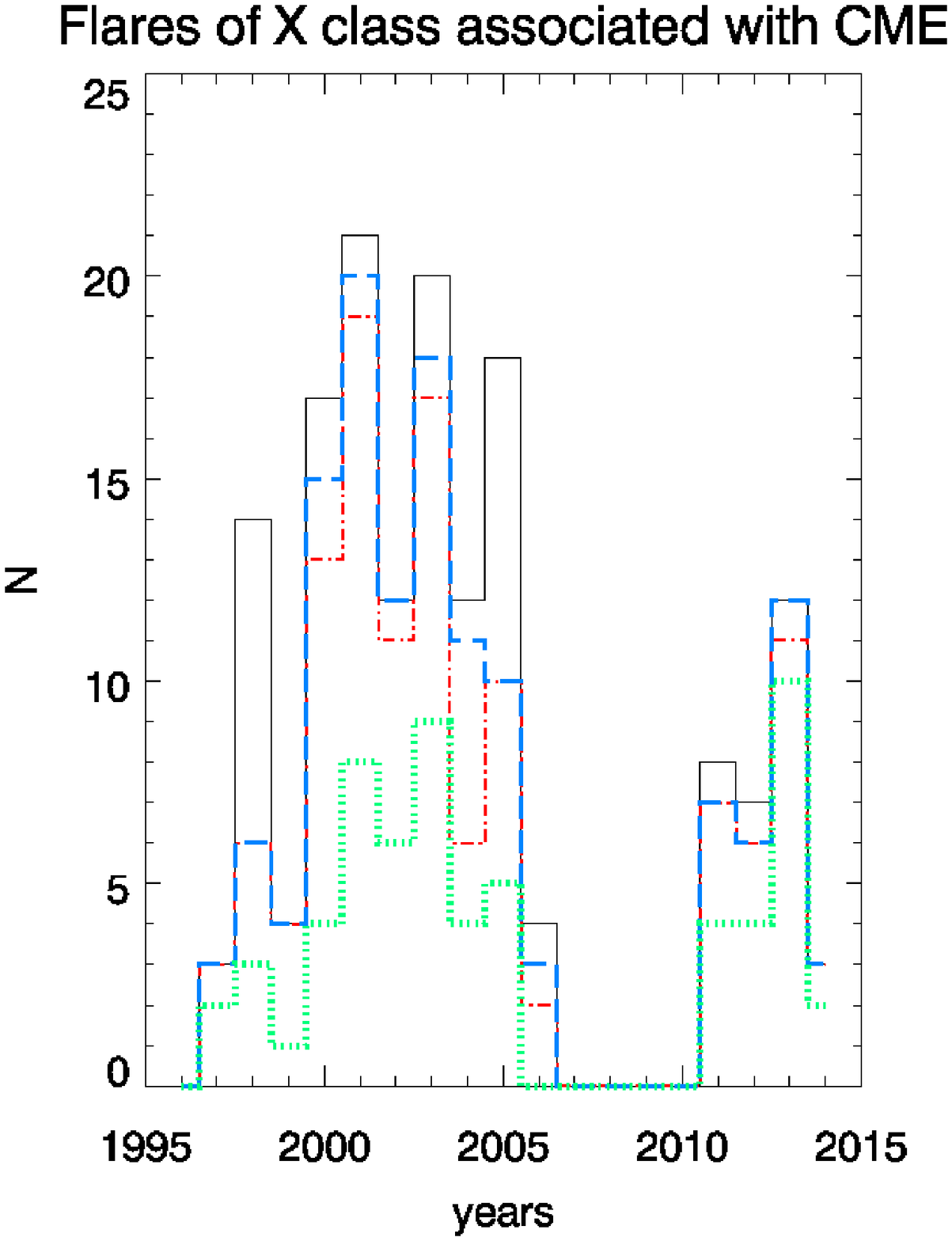}\\

\caption{Total distribution of C-class ( top panel), M-class ( middle panel),and X-class ( bottom panel) flares associated with the CMEs as a function of years in $\pm$ two hour time window (long--dashed-blue line), in $\pm 1$ hour time window (dot--dashed-red line) and $\pm 30$ minute time window (dotted-green line) for CDAW. The black line in each panel indicates the distribution of all flares for a given class.}
\label{fig10}
\end{figure*}

\begin{figure*}
\centering
\includegraphics[width=0.45\textwidth]{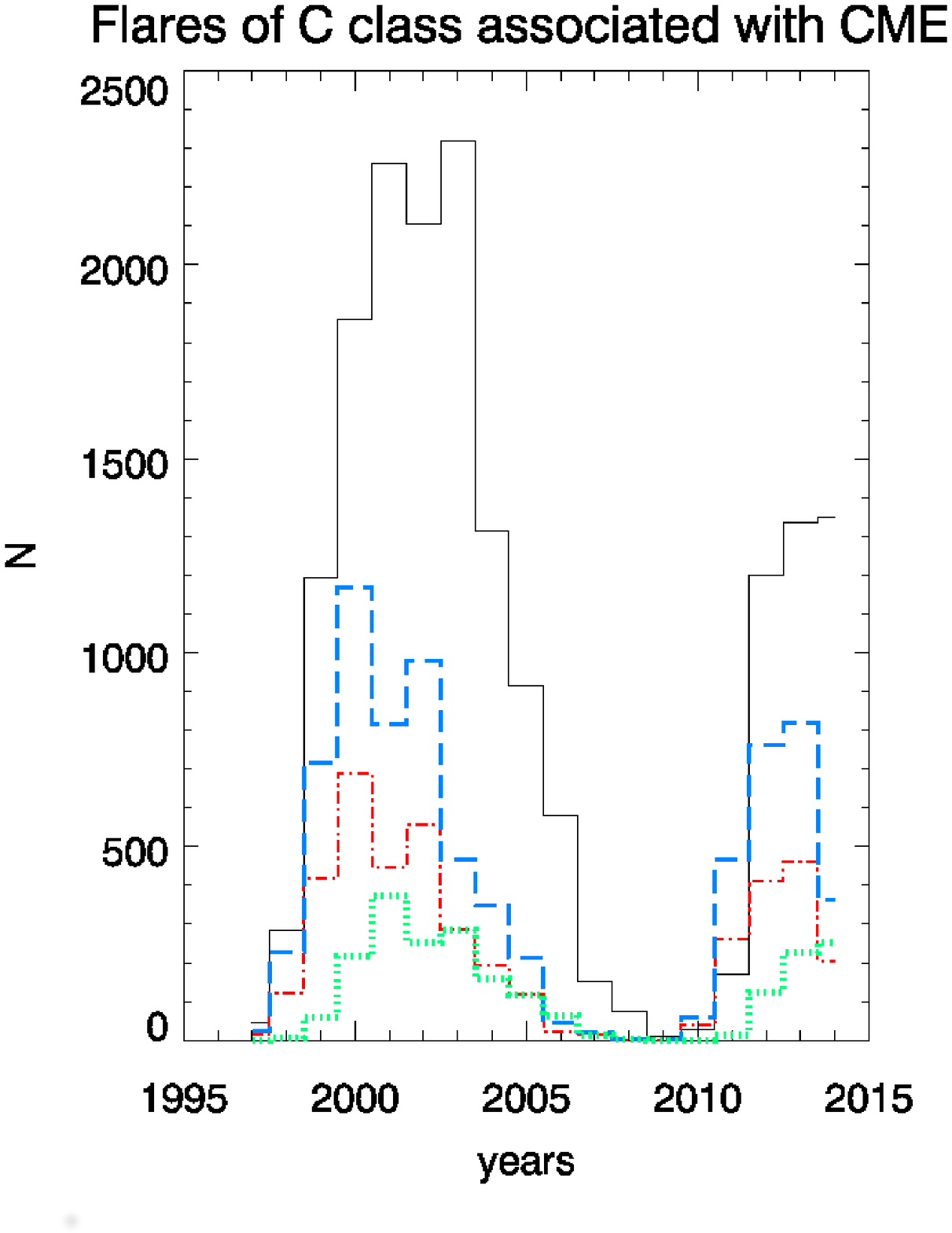} \\
\includegraphics[width=0.45\textwidth]{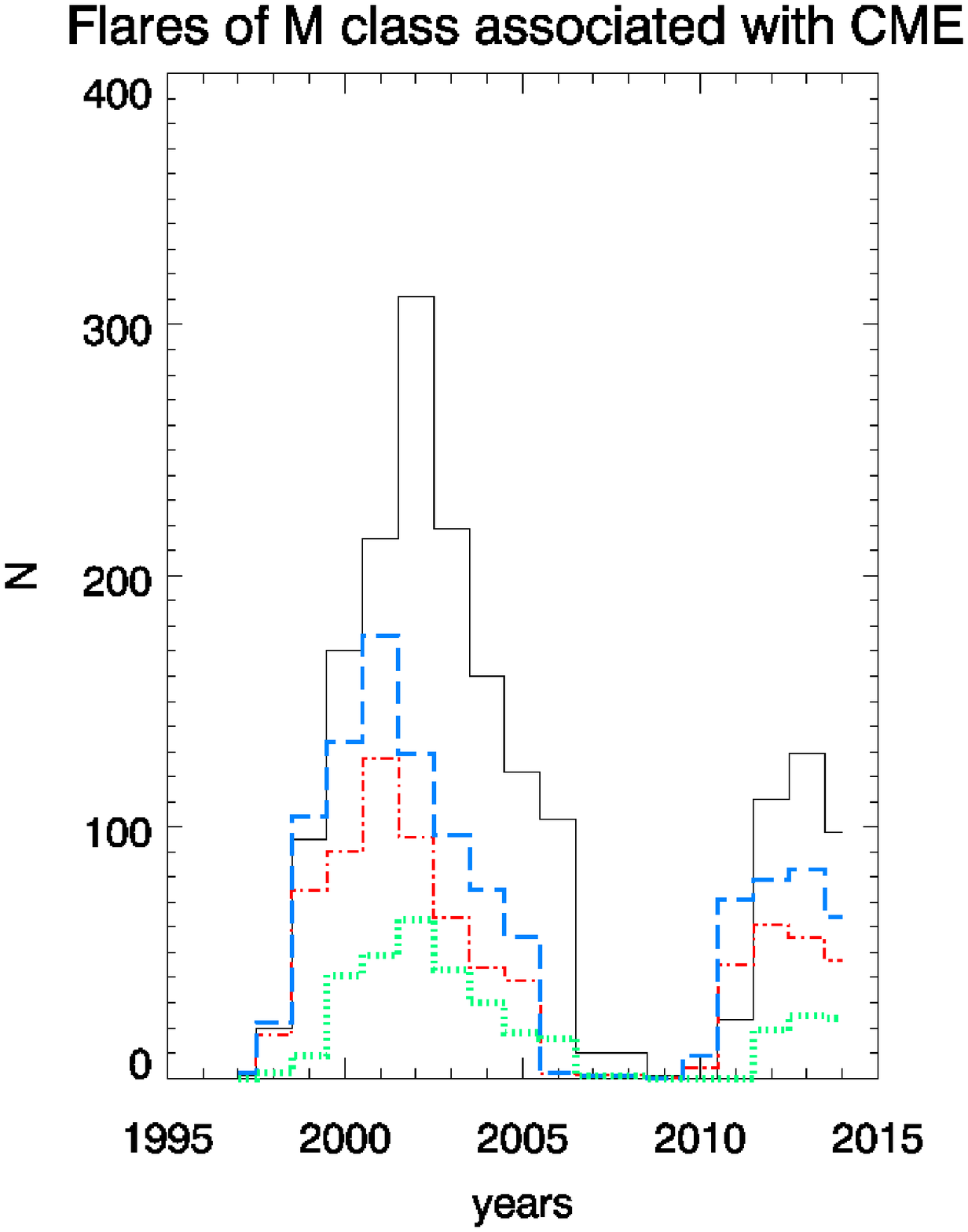}\\
\includegraphics[width=0.45\textwidth]{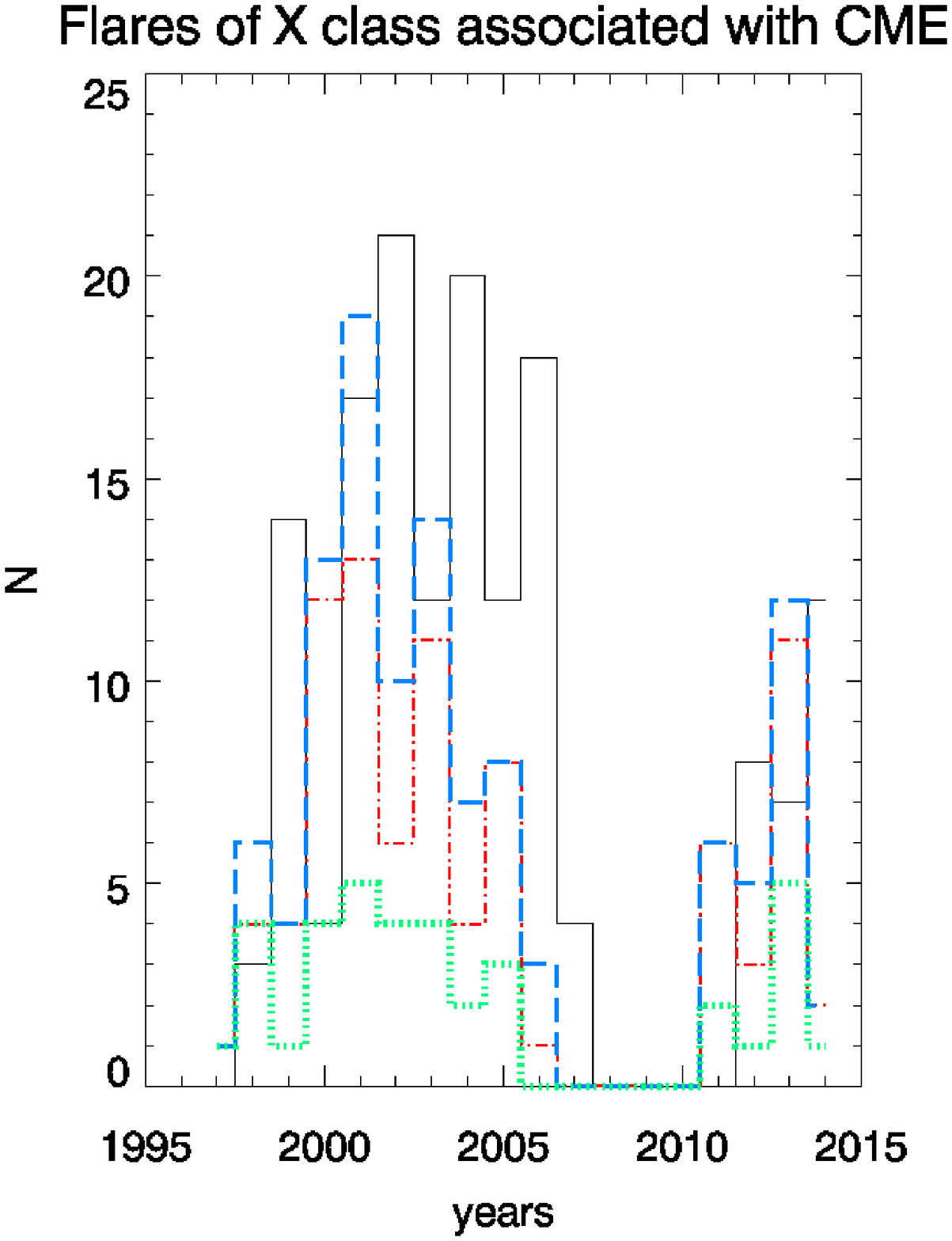}\\

\caption{Total distribution of C-class ( top panel), M-class ( middle panel),and X-class ( bottom panel) flares associated with the CMEs as a function of years in $\pm$ two hour time window (long--dashed blue line), in $\pm 1$ hour time window (dot--dashed-red line) and $\pm 30$ minutes time window (dotted-green line) for CACTus. The black line in each panel indicates the distribution of all flares for a given class.}
\label{fig10b}
\end{figure*}

The distributions  of C-,  M-, and X-class flares associated with CMEs, as a function of the year in the solar cycle in the time interval of $\pm$ two hours, $\pm$ one hour, $\pm 30 $ minutes, are shown in Figures \ref{fig10} and \ref{fig10b} for CDAW and CACTUs, respectively. We note that the probability of finding CMEs associated with flares decreases when the time window is narrower. However, we see that the peak and the shape of the distributions remain similar, independently of the temporal time window considered. In the (top panel) of Figures \ref{fig10} and \ref{fig10b}, the peaks correspond to the years 2000\,--\,2002 and 2011\,--\,2014, in agreement with the solar-activity cycles (see Figure \ref{fig1}). We can see in the distribution of C-class flares associated with CMEs (top panels of Figures \ref{fig10} and \ref{fig10b}) that these events are  present also during the phases of minimum of solar-activity. In the distribution of M-class flares( middle panels of Figures \ref{fig10} and \ref{fig10b}) we can see a trend very similar to the one found for C-class flares.
On the other hand, we note that the distribution of X-class flares associated with CMEs is more uniform across the solar cycle than that of C- and M-class flares for both datasets. For example, for CDAW in 1998 and 2005 we observe 14 and 18 X-class flares associated with CMEs respectively, while for CACTus we observe  6 and 8 X-class flares in those years, although the magnetic activity was not high. However, when we consider only the flares associated with CMEs in a $\pm 30 $ minute time window, we find a distribution of the X-class flares more consistent with the solar cycle for both distributions

\begin{figure*}[h!]
\centering
\includegraphics[width=0.35\textwidth,trim=50 130 50 130]{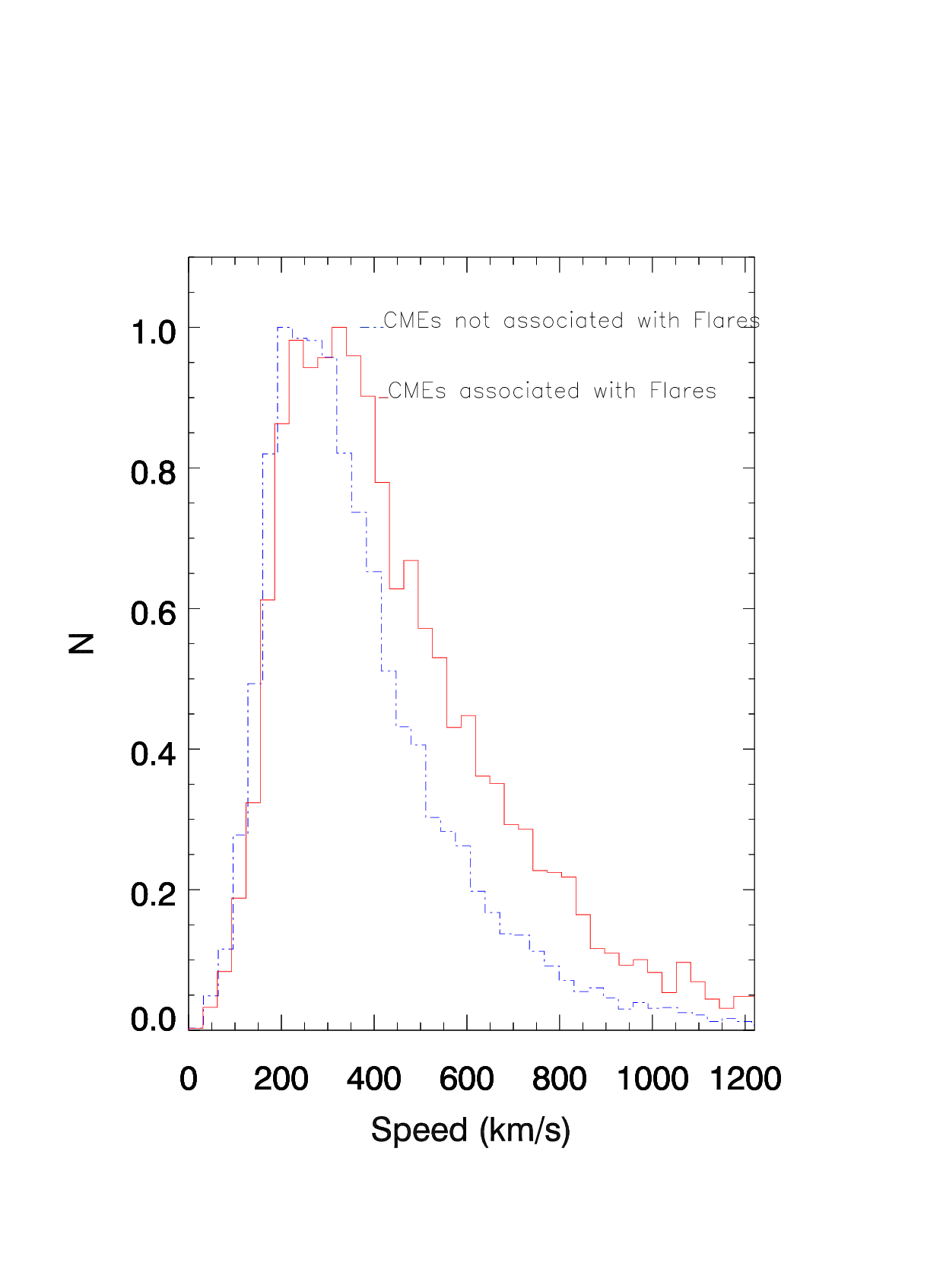} 
\includegraphics[width=0.35\textwidth,trim=50 120 50 120]{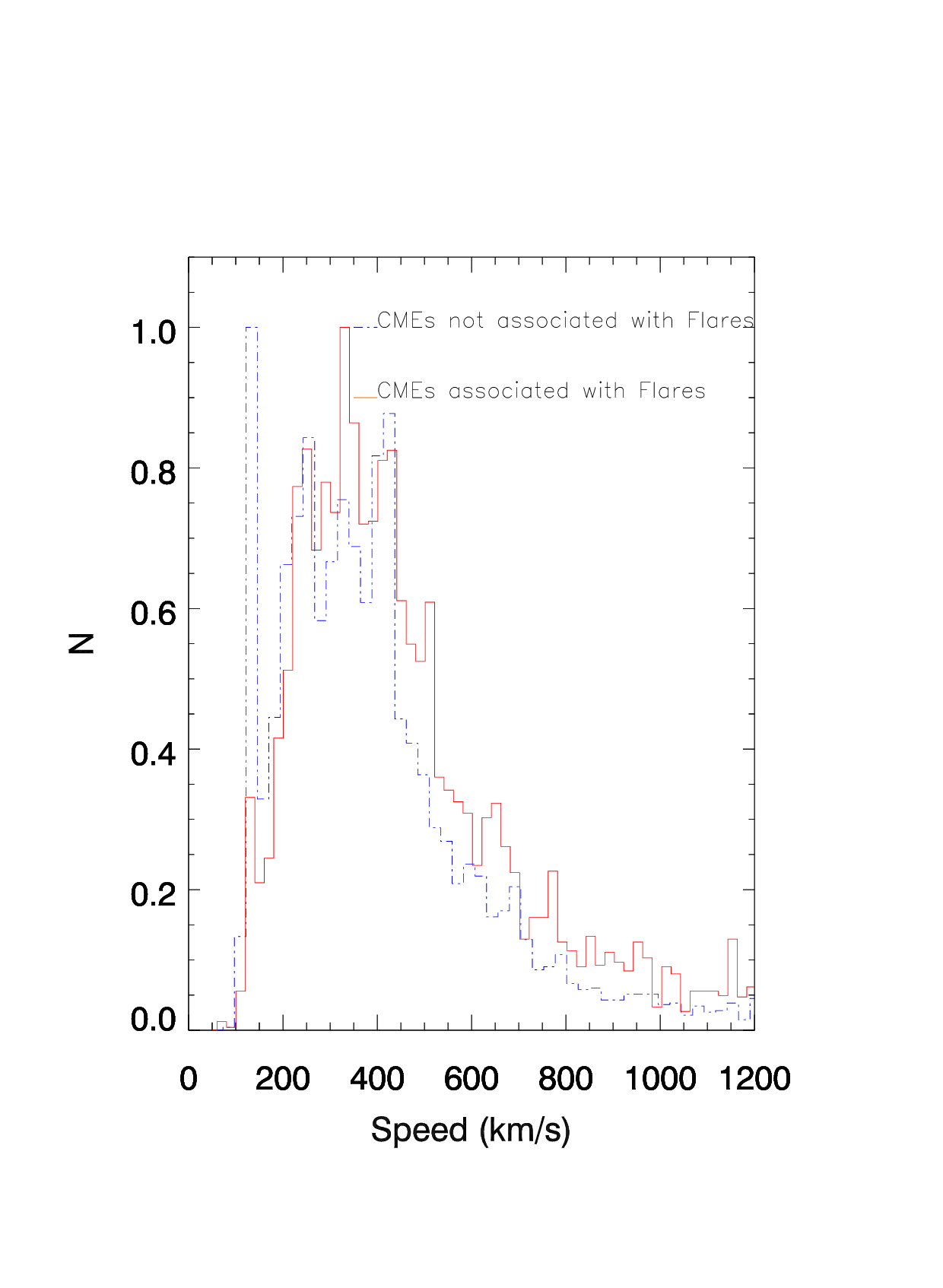}
\caption{Distribution of the CME velocities for CDAW (left panel) and CACTus (right panel) datasets. The solid-red and dot--dashed-blue lines correspond to CMEs associated with flares and not associated with flares, respectively.}
\label{fig11}
\end{figure*}

In Figure \ref{fig11} we show the distribution of CMEs  velocity, distinguishing between events associated with  flares in the $\pm$ two hours time window and events not associated with flares.  The mean CMEs velocities in CDAW datasets for the CMEs associated and not associated with flares are 472.87 $\pm$ 2.77 km\,s$^{-1}$ and 379.41 $\pm$ 2.25 km\,s$^{-1}$, respectively. The mean CMEs velocities in CACTus dataset for the CMEs associated and not associated with flares  are  500.62 $\pm$ 3.28 km\,s$^{-1}$ and 437.75 $\pm$ 3.79 km\,s$^{-1}$ respectively. In Figure \ref{fig11} (right panel) we note that there are a great number of  CMEs not associated with flares in the CACTus dataset with velocities  between 100\,--\,200  km\,s$^{-1}$.  

\begin{figure*}[h!]
\centering
\includegraphics[width=0.40\textwidth,trim=50 120 50 120]{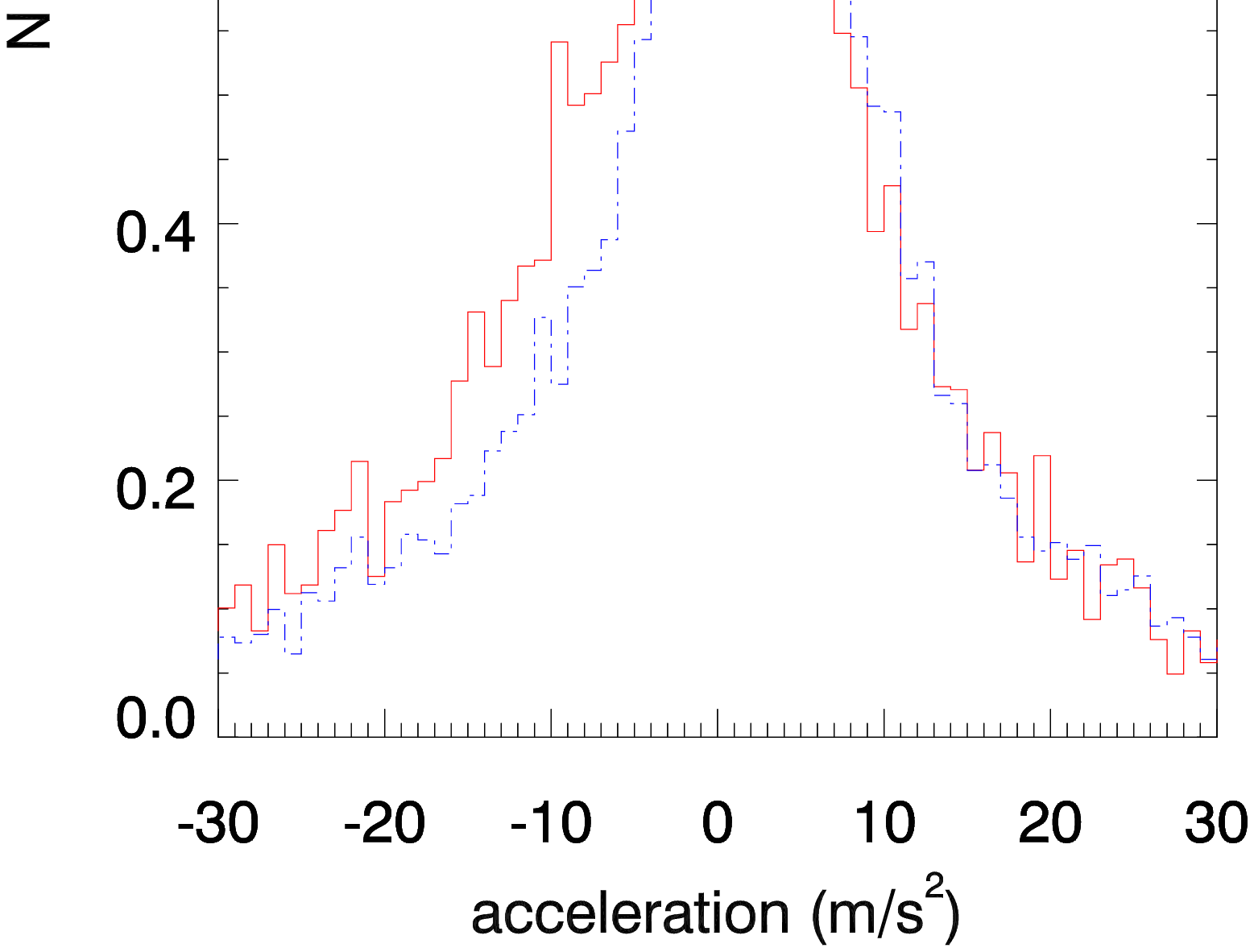}
\includegraphics[width=0.40\textwidth,trim=50 120 50 120]{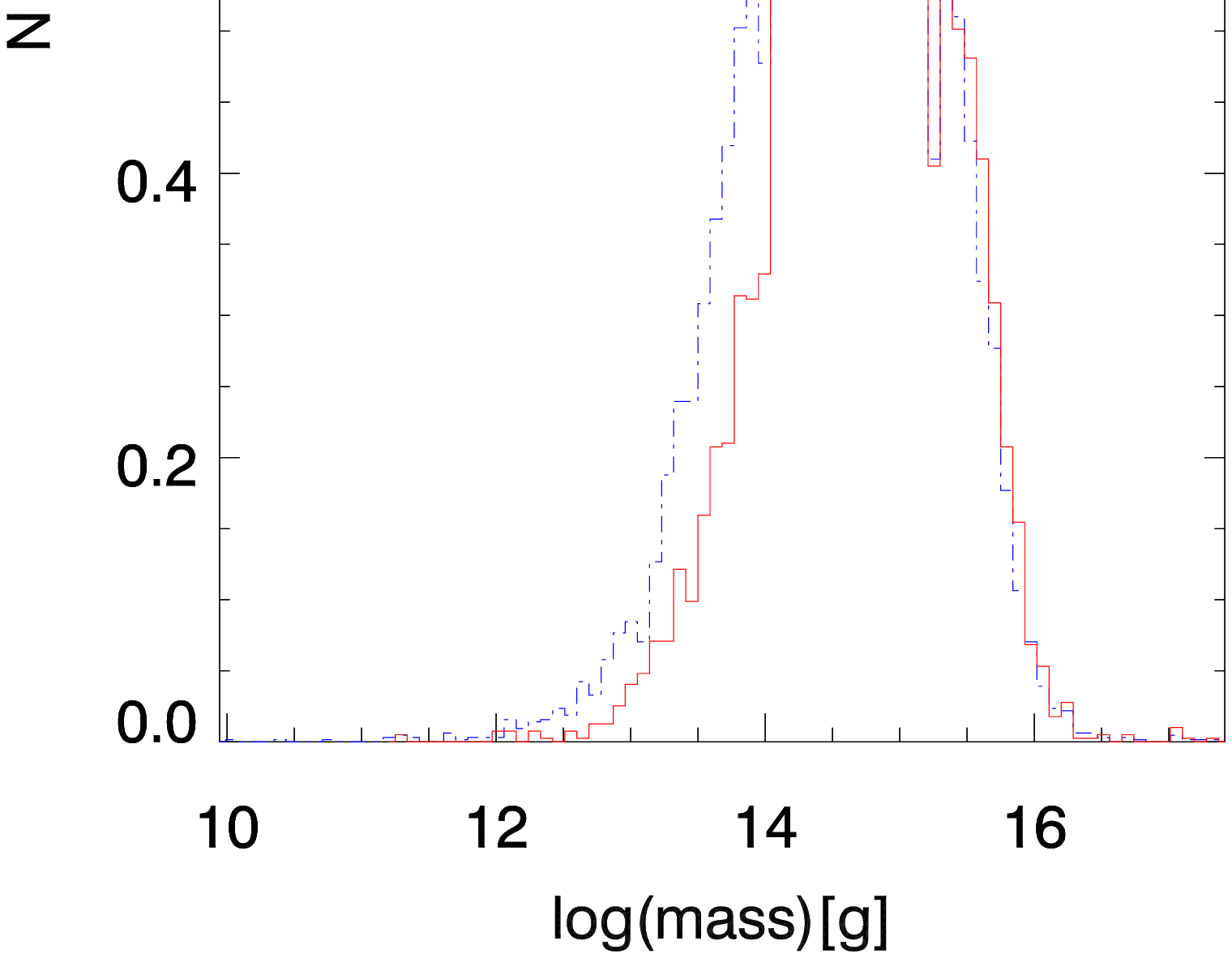}

\caption{Distribution of the CME acceleration (left panel) and mass (right panel) for CDAW dataset. The solid-red and dot--dashed blue-lines correspond to CMEs associated with flares and not associated with flares, respectively.}
\label{fig11d}
\end{figure*}

The distribution of the CME acceleration ( left panel of Figure \ref{fig11d}) shows that the CMEs associated with flares have an average acceleration  of -$0.32 \pm$ 0.34 m\,s$^{-2}$ , while the CMEs not associated with flares have an average acceleration of $3.44 \pm 0.39$ m\,s$^{-2}$ .
In the (right panel of Figure \ref{fig11d}) we report the distribution of the CME mass.
The logarithms of mean CME mass [g] for CMEs associated with flares and not are $14.70 \pm 0.006 $ g and $14.54 \pm 0.004$, respectively.

\begin{figure*}[h!]
\centering
\includegraphics[width=0.48\textwidth,trim=50 120 50 120]{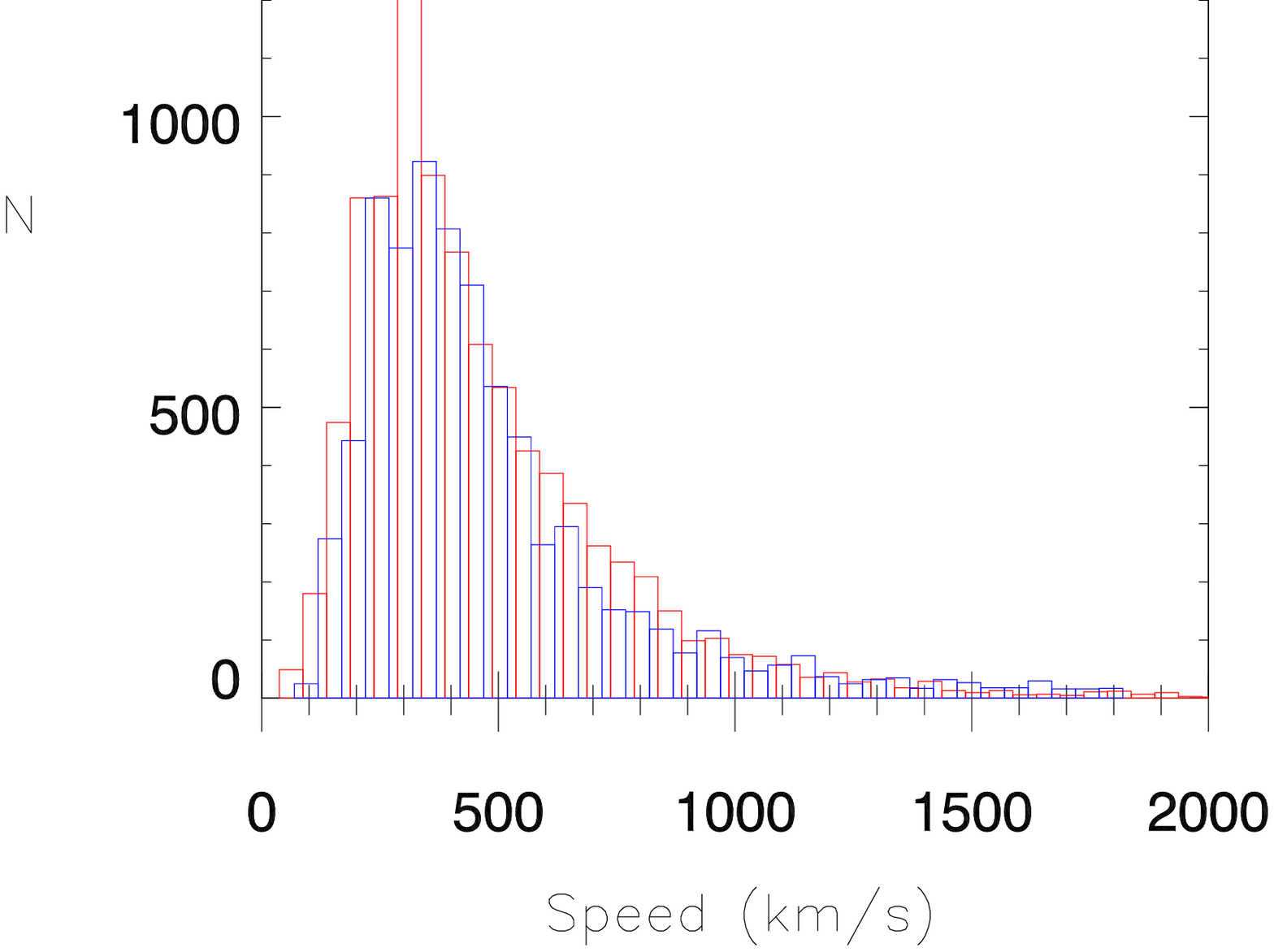} \\
\includegraphics[width=0.48\textwidth,trim=50 120 50 120]{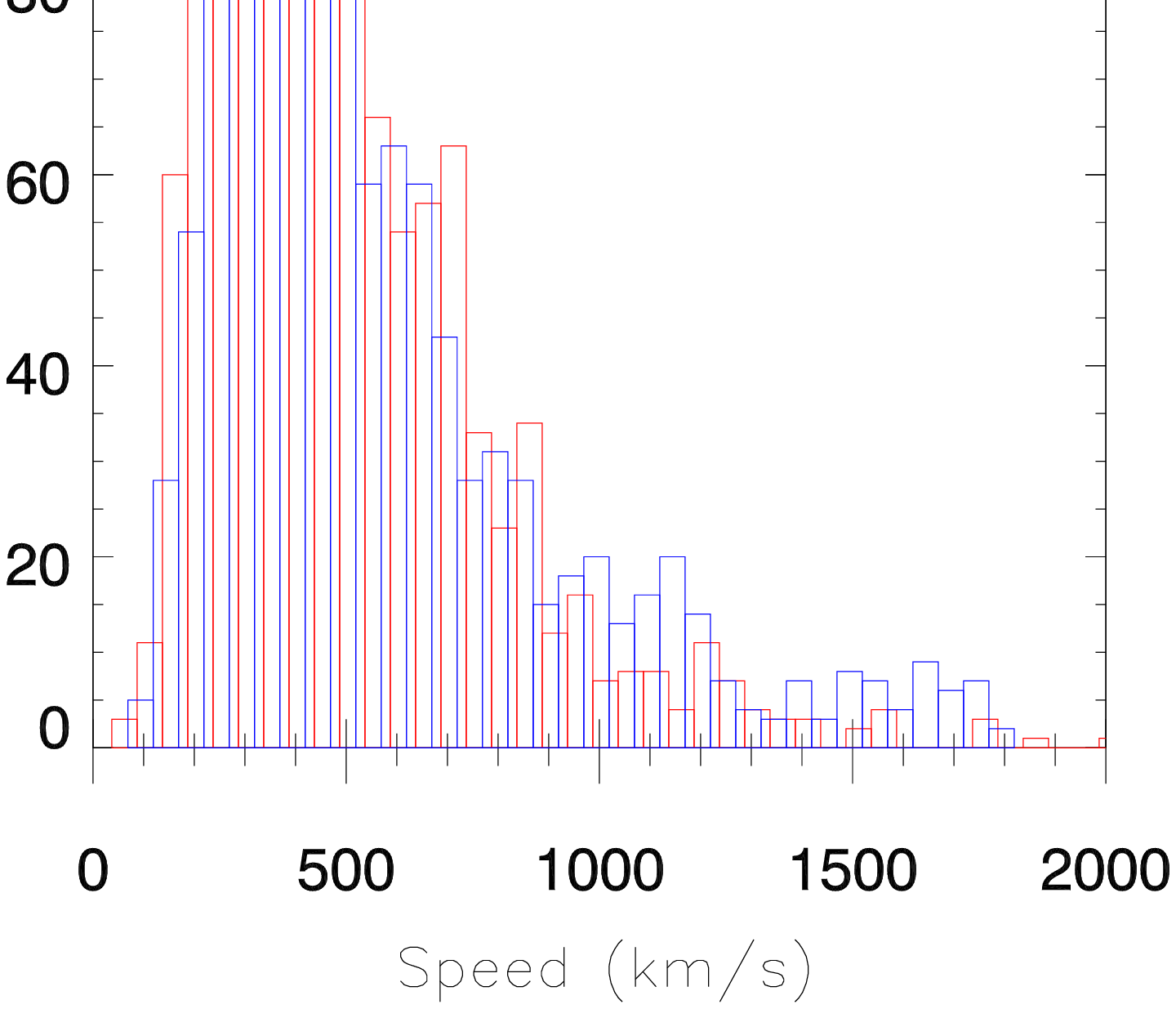}\\
\includegraphics[width=0.48\textwidth,trim=50 120 50 120]{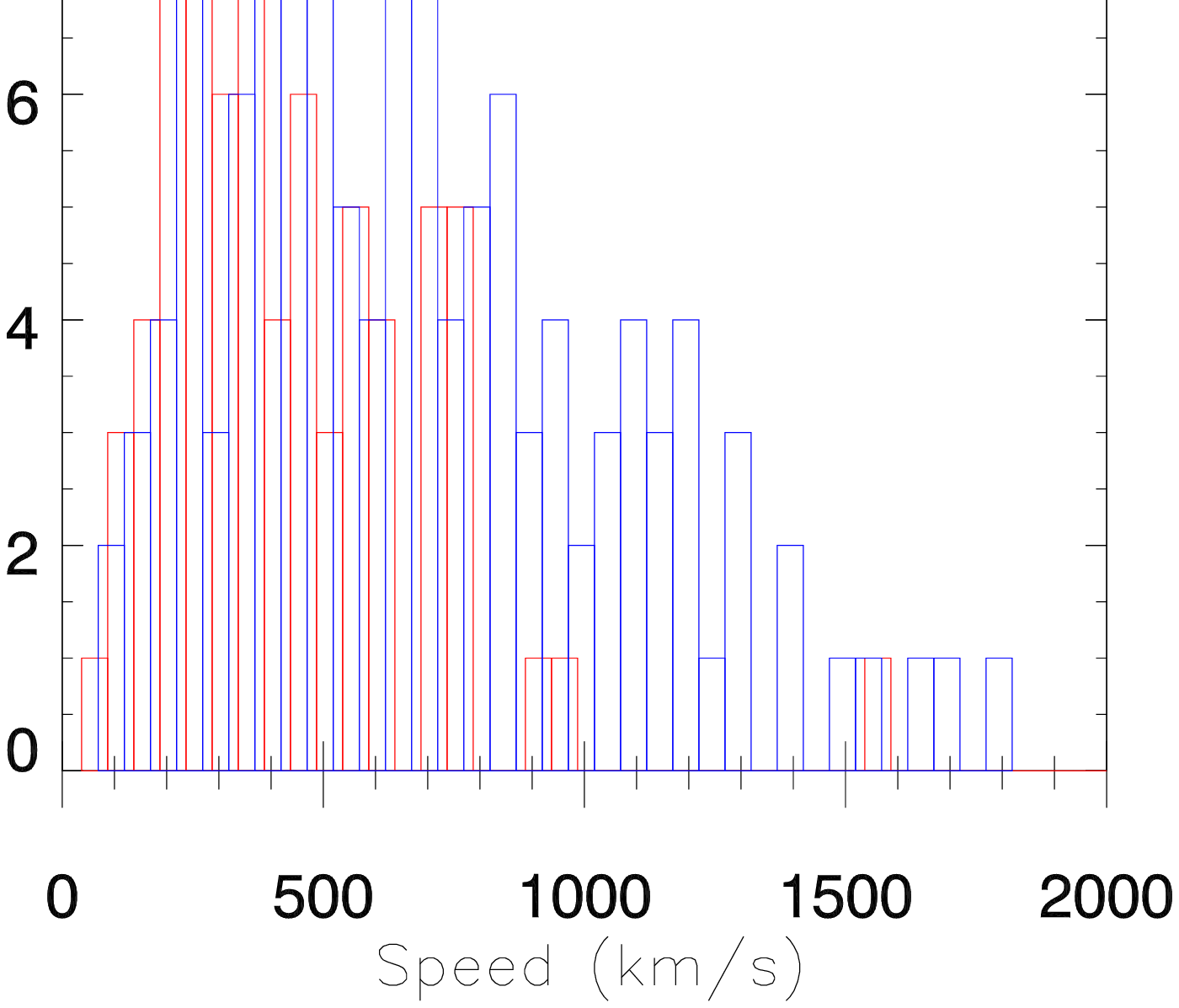}\\
\caption{Distribution of the linear velocity of CMEs associated with flares of C-class (top panel), M-class ( middle panel), and X-class (bottom panel)  For CDAW (red line) and CACTus (blue line)}.
\label{fig12}
\end{figure*}

Further we found different linear velocity distributions when we distinguish among the three flare classes (Figure \ref{fig12}). The CDAW velocity distribution of the CMEs associated with the C-class flares shows a significant peak at 250 km\,s$^{-1}$ that is not present in the CACTus velocity distribution. For CDAW the mean linear velocities of CMEs associated with flares of C-, M-, and X-classes are $535.95 \pm 1.11$ km\,s$^{-1}$ , $585.71 \pm$ 8.16 km\,s$^{-1}$, and $626.99 \pm$ 56.58 km\,s$^{-1}$, respectively. For CACTus the mean velocities of CMEs associated with flares of C-, M-, and X-classes  are $487.91 \pm$ 3.46 km\,s$^{-1}$, $562.16 \pm$ 9.82 km\,s$^{-1}$, and $688.15 \pm$ 34.35 km\,s$^{-1}$ , respectively. 

 Both distributions are similar except that CACTus finds fewer events than CDAW, as we note in the distribution of CMEs associated with C-class flares (see the blue line in Figure \ref{fig12}). It is worth noting that the differences between the velocity of CMEs associated with X-class flares in the CDAW dataset and in the CACTus dataset. In fact the CMEs associated with flares in the CACTus dataset are faster than the ones of the CDAW dataset. 

\begin{figure*}[h!]
\centering
\includegraphics[width=0.50\textwidth]{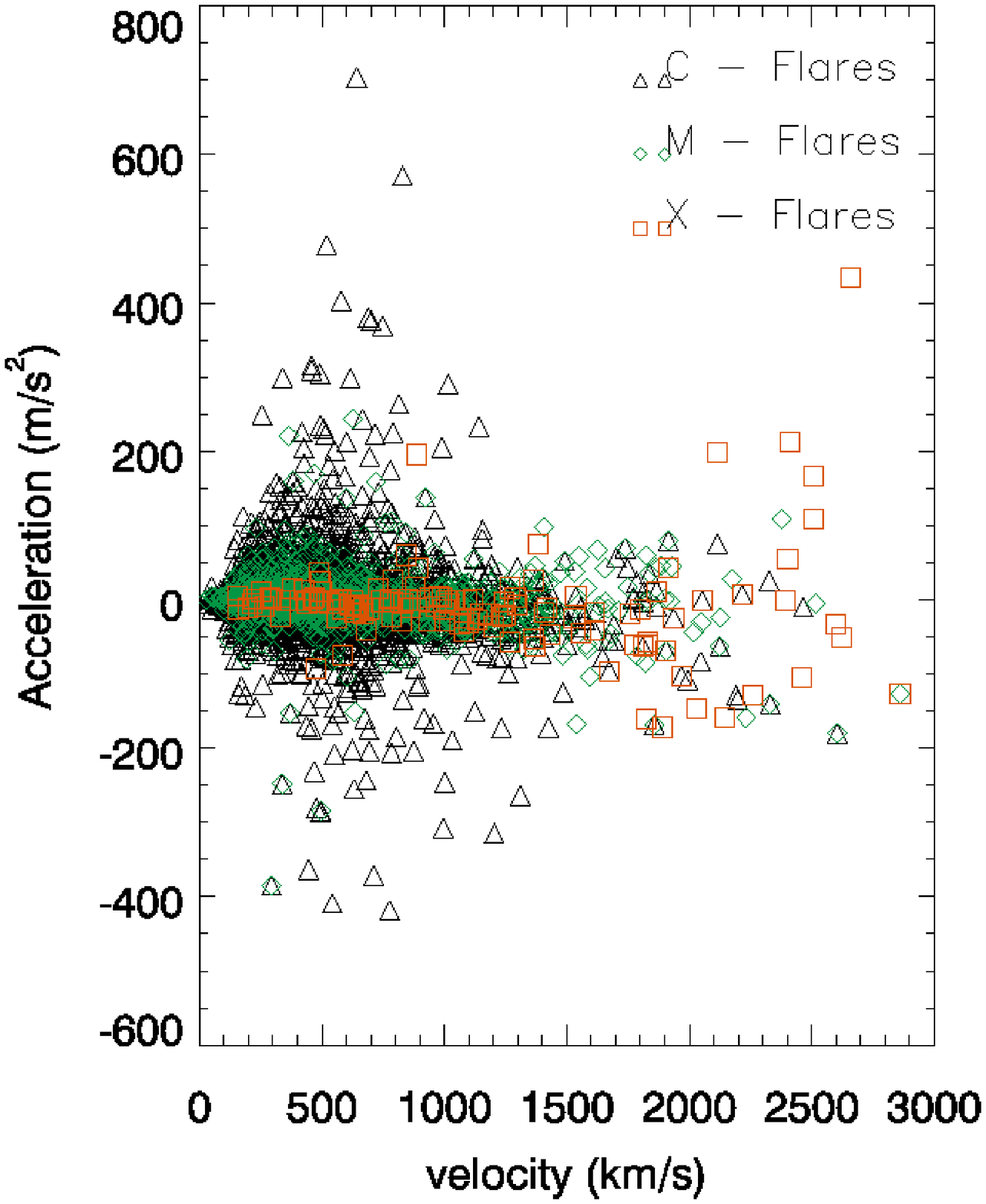}
\caption{Scatter plot of the CME velocity as a function of their acceleration. We distinguished the associated flares of C-class (black triangle), M-class (green diamonds), and X-class (red square).}
\label{fig13}
\end{figure*}

In Figure \ref{fig13}  we show the relationship between CME speed and acceleration, taking into account the different class of the associated flares. We found that the majority of CMEs characterized by higher velocities are associated with acceleration between $-200 $ and 200 m\,s$^{-2}$ . On the other hand, there are many CMEs with a wide range of acceleration (between $-400 $ and 400 m\,s$^{-2}$) but with velocity below 700 km\,s$^{-1}$. These CMEs are mainly associated with C- and M-class flares.

 \section{CME Parameters and Flare Energy} 
      \label{S-CME parameters and flare energy} 

\begin{figure*}[h!]
\centering
\includegraphics[width=0.50\textwidth]{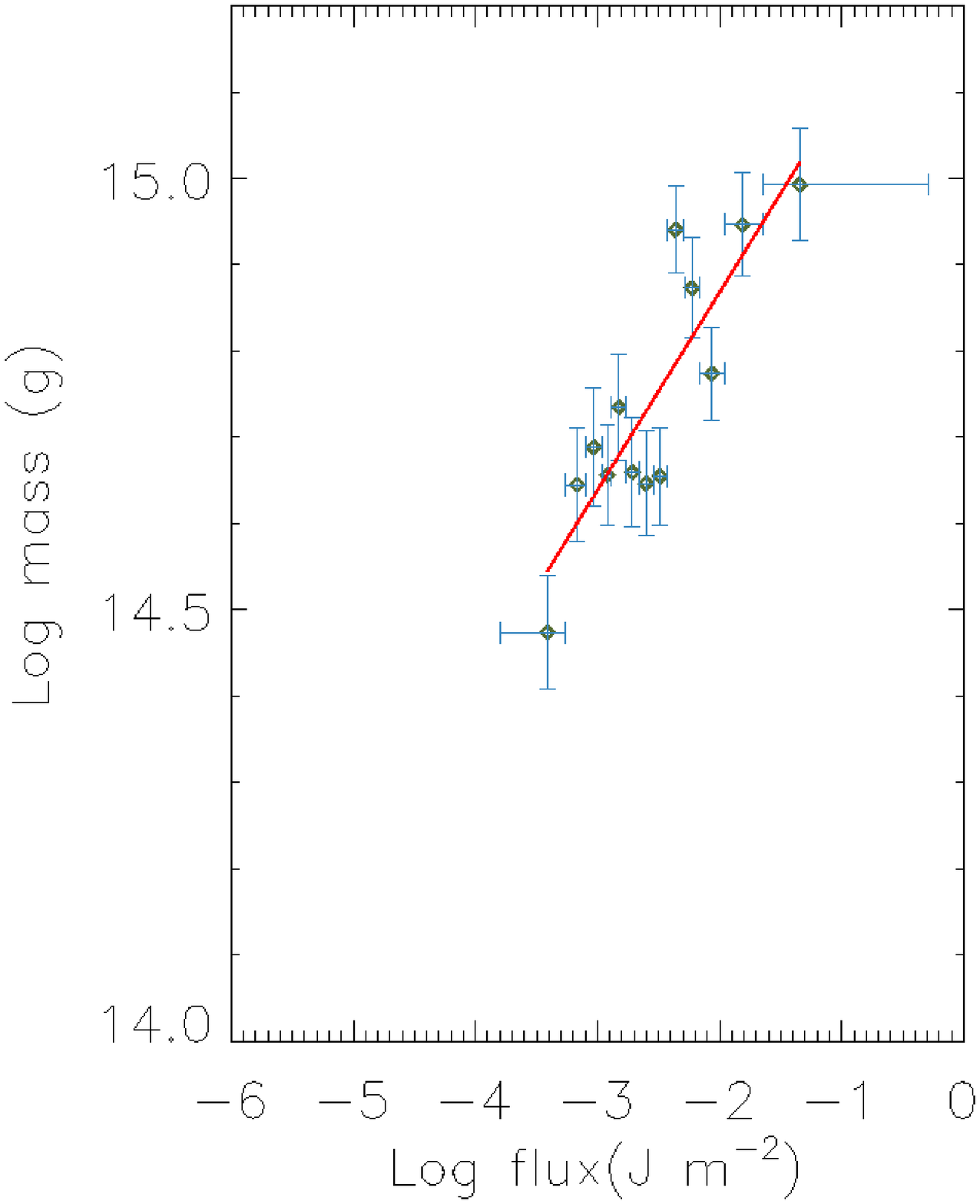}
\caption{{ Relationship between integrated flare flux and CME mass for the CMEs associated with flares in the $\pm$ two hour time window.}}
\label{fig14}
\end{figure*}

We used our dataset to investigate the relationship between the logarithm of the flare flux and the logarithm of the CME mass. In this case we considered not only the temporal correlation between flares and CMEs, but also their spatial correlation. We limited this analysis to flares which occurred in the time window of $\pm$ two hours and characterized by a  known location of the source region on the solar disc. We associated flares which occurred in the top-left quadrant of the solar disc with CMEs characterized by a polar angle between 0$^{\circ}$ and 89$^{\circ}$, flares occurred in the bottom-left quadrant with CMEs characterized by a polar angle between 90$^{\circ}$ and 179$^{\circ}$, \ecc This criterion of association between flares and CMEs is based on the assumption that most CMEs propagate nearly radially, according to the standard CME--flare model (see \inlinecite{Lin00}). 
In this way we obtained  1277 CMEs that are spatially and temporally correlated with flares. We considered the flux of these flares integrated from their start to end in the 0.1\,--\,0.8 nm range. Subsequently the 1277 CME--flare pairs were binned into 13 equal sets of 100 pairs each, with the exception of the last one containing 77 pairs. We computed the mean value of the logarithm of the CME mass  and  the logarithm of flare flux in each group. The relationship between the logarithm of the CME mass and the logarithm of the flare -flux is shown in Figure \ref{fig14}, where we also show the error bars of the logarithm of the CMEs mass computed as $\sigma/\sqrt{N}$. The flux error bars correspond to the minimum/maximum flare flux values spanned by that bin. The results shown in Figure \ref{fig14} are similar to the result reported by \cite{aar2011}. In fact we found the following log--linear relationship between integrated flare flux [$\phi_f$] and CME mass [$m_{CME}$]:
 \begin{equation}
 \log(m_{CME})=(15.33 \pm 0.10) + (0.23 \pm 0.04) \,\log(\phi_f).
\end{equation}
However, it worth noting that this relationship disappears when we limit the same analysis  to each year of our dataset (we do not show the corresponding plots in this article). This means that the log--linear relationship between integrated flare flux and CME mass is consistent only for a large sample of events, but it is not valid during the different phases of the solar cycle. 
 
The unreduced $\chi^{2}$-test of the linear fit that we have performed is 0.07. Further tests to verify the goodness of the fit have also been performed. The Spearman's coefficient , indicating how well the relationship between two variables can be described using a monotonic function, is 0.82 which indicates a good correlation between these quantities. We also found a p-value of 0.85 and a linear Pearson correlation coefficient of 0.87.

\section{Conclusions} 
      \label{S-conclusions}

In this article we used the huge dataset of  CMEs observed for nearly all of the operational time,to date, of the LASCO mission onboard  \soho to infer some properties of these events over Solar Cycles 23 and 24 and to study their correlation with the flares observed in the X-ray range between 1.0 and 8.0 $\AA$, by GOES.  We used two CME catalogs; one based on manual identification  of CMEs (CDAW) and one based on their automatic tracking (CACTus).   

 From the analysis that we have performed we conclude that the peak in the number of CMEs in Cycle 24 is higher than the number during Cycle 23.  In particular, for the CDAW dataset the number of CMEs at the  maximum of Solar Cycle 24 is higher than the number at the maximum of  Solar Cycle 23; for CACTus the peaks during the maxima are similar, but the one corresponding to Solar Cycle 24 is more extended in time than the one corresponding to Solar Cycle 23.  This result seem to be in contrast with the fact that the magnetic activity during  Cycle 24 was weaker than during  Cycle 23, as noted in the work of \cite{tjh2015} who analyzed the frequency shifts of the acoustic solar-modes measurement separately for the two cycles and found that the magnetic activity during Solar Cycle 24 was weaker than during Cycle 23.\\ Solar Cycle 24 has been extremely weak as measured by the sunspot number (SSN) and is the
smallest since the beginning of the space age. The weak activity has been thought to be due to
the weak polar field strength in Cycle 23. Several authors have suggested that the decline in cycle
24 activity might lead to a global minimum. (\citealp{pad2015}; \citealp{zol2014}).\\ 
From the analysis of the CME average velocity, we found two peaks that  reflect the solar-activity cycles and can be interpreted, according to \cite{qiu2005} as an effect of the magnetic flux involved by the events during the solar maxima, but we also observe another peak at the minimum of the cycles, even if only for the CACTus dataset, in the year 2009. This peak agrees with the cyclic variation of the CME velocities in the previous solar-activity cycles, as reported by \cite{Ivanov2001}. 

From the distribution of the average acceleration of the CMEs, we see a maximum of about 15 m\,s$^{-2}$ $\pm$ 2.71 m\,s$^{-2}$ , corresponding to the minimum of the distribution of the average velocity (see Figure \ref{fig2}, upper and middle panel). We think that this peak may be due to the contribution of the slower CMEs occurring  during solar-activity minimum;  These CMEs are characterized by higher positive values of acceleration. 

We found also that the tail of the distribution of the higher velocities of the CMEs observed during the descending part of Solar Cycle 23 (from 2000 to 2006) is no longer present when going from the maximum (2000) to the minimum (2006) of the cycle (see Figure \ref{fig3}).
   
 The distribution of the CME angular widths for CDAW shows that on average, the narrower CMEs are slower and the majority of the CMEs are characterized by an angular width lower than $100^{\circ}.$ Only during the maximum of the  solar cycle (2000 and 2001) do we observe a significant number of CMEs with an angular width larger than $100 \pm 0.44^{\circ} $  

The CDAW and CACTus datasets present a different amplitude of the range spanned by the mean angular width, \ie for the CACTus catalog the mean width varies from $\approx 30^{\circ}$ during solar-activity minimum  to $\approx 40^{\circ}$ during the maximum of activity, while for CDAW the mean width varies from $\approx 20^{\circ}$ to $\approx 80^{\circ}$.

The latitude distribution of CMEs follows the latitude distribution of the closed-magnetic-field regions of the corona, which is consistent with the fact that CMEs originate in closed-field regions \citep{hun93}. We also note that the distribution of PA changes in time from a broader distribution in 2000 (near the maximum of Solar Cycle 23) to a more peaked distribution in 2006 (near the minimum of solar activity).

Using the dataset of CMEs and flares and selecting the event occurring in the same time window of $\pm 2$ hour,$\pm 1$ hour and $\pm 30 $ minutes we identified CMEs and flares that are temporally correlated. 
 
Although the number of associated CMEs--flares that are both temporally and spatially correlated might seem low, \cite{aar2011}, studying the correlation between flares flux and CMEs mass, found a similar result with 826 associated CMEs--flares during the time interval 1996\, -- \,2006.
\cite{you2012}, studied the correlation between flare flux and CME energy, and found 776 associated CME--flares during the time interval 1996\,--\,2010. 
Considering the flare start time, we found that the highest number of CMEs and flares  detected in the CDAW dataset (59.57\,\%) are characterized by a difference in time between 10 and 80 minutes (see the black line in the top panel of Figure \ref{fig9}), in agreement with \citet{aar2011}. 
The CME--flare associated events for the CACTus dataset show a wider temporal range. We argue that this difference between the two datasets depends on the different criteria used by the observer for defining a CME in CDAW. A time window of 10--80 minutes is clear evidence that in many cases the flare occurs before the first observation of the CME in the coronagraph and, taking into account that the temporal resolution of LASCO is about 30 minutes, 
There are a number of cases in which the flare most likely precedes the CME initiation and may be the first manifestation of the initiation process. One should also need to investigate the role of filaments/prominences in the initiation.

The shape of the distributions of C-, M-, and X-class flares associated with CMEs varies with the intensity of the flares, (see Figures  \ref{fig10} and \ref{fig10b}), but it is similar for both datasets. In particular, we note that the distribution of X-class flares associated with CMEs is quite uniform with respect to the one of C- and M- class flares across the solar cycle. However, when we consider only the flares associated with CMEs in the $\pm 30 $ minutes time window, we find a distribution of the X-class flares more consistent with the solar cycle (see Figure \ref{fig10}, bottom panel).

We found that most of the CMEs characterized by higher linear velocities are associated with flares (see-\eg -  \cite{gos1976}; \cite{mon2003}).  
  The mean velocities for CMEs associated  with flares are higher than the velocities for CMEs not associated  with flares in both datasets. Our results are therefore very similar to those found by \cite{aar2011}.  

Moreover, our analysis  shows that the width of the CMEs associated with flares is positively correlated with the flare flux.   The mean angular width of the flares associated with CMEs is  $68.61 ^{\circ}$, $116.82 ^{\circ} $, $258.49 ^{\circ}$ for C-, M-, and X-class flares, respectively. 

 The distribution of the CME acceleration ( left panel of Figure \ref{fig11d}) shows that the CMEs  associated with flares have an  average acceleration of -0.32 $\pm$ 0.34 m\,s$^{-2}$ , while the CMEs not associated with flares have an average positive acceleration of 3.44 $\pm$ 0.37 m\,s$^{-2}$ . We suggest that these values are slightly different from those found by \cite{aar2011} due to the diverse sample of events considered in our article.
 
We also used our dataset to further extend the study on the log--linear relationship between the flare flux [$\phi_f$] and the CME mass [$m_{CME}$] performed by \citet{aar2011}. In this case, we considered not only the temporal correlation between flares and CMEs, but also their spatial correlation. As mentioned above, this allows us to be more confident that the CMEs and the flares may be linked, although some events may be neglected. We found that $\log(m_{CME}) \propto 0.23\,\log (\phi_f)$.
The differences between the results of \cite{aar2011} and ours may be due to several reasons. First of all we considered a more extended dataset. We also used the flux of the flares integrated from their start to end in the 0.1\,--\,0.8\,nm range. Finally, we used different criteria to determine the association between flares and CMEs.  Therefore, we conclude that the log--linear relationship is valid not only when we consider the peak of the flare flux, but also when we consider the energy released during the whole event.        
 
It is worth of note that this relationship disappears when we limit the sample of flare--CME pairs to the different phases of the solar cycle. This means that the log--linear relationship is valid only from a statistical point of view, \ie when we consider a large sample   of events.
 
We argue that this result is due to  different aspects (intensity of magnetic field, magnetic reconnection, configuration of sunspots on the solar surface) that influence the evolution of these phenomena. Further study and analysis must be made on  the intensity of the magnetic flux involved in these phenomena and consequently on the capacity to eject the mass into in the interplanetary space. In fact, the magnetic configuration can play an important role in determining the different ways to buildup and release magnetic free energy and the role of magnetic reconnection


\begin{ack}

The authors wish to thank the anonymous referee for their useful suggestions which allowed to improve the article.
This research work has received funding from the European Commission’s Seventh Framework Programme under the Grant Agreements no. 606862 (F-Chroma project) and no. 312495 (SOLARNET project). This research is also supported by the ITA MIUR-PRIN grant on ``The active sun and its effects on space and Earth climate'' and by Space WEather Italian COmmunity (SWICO) Research Program. We are also grateful to the University of Catania for providing the Grant FIR 2014.

\end{ack}

\section{Disclosure of Potential Conflicts of Interest}
The authors declare that they have no conflicts of interest.

\end{article}

\end{document}